\documentclass[apj]{emulateapj-rtx4}
\usepackage{natbib}

\newcommand{\bptx}{[NII]6584/H$\alpha$}
\newcommand{\bpty}{[OIII]5007/H$\beta$}

\makeatletter

\newenvironment{inlinefigure}{%
\def\@captype{figure}%
\noindent\begin{minipage}{0.999\linewidth}\begin{center}}
{\end{center}\end{minipage}\smallskip}
\makeatother

\shorttitle{Luminosity-dependent Emission-line Diagnostics}
\shortauthors{Cowie, Barger \& Songaila}

\begin{document}

\title{Luminosity dependence and redshift evolution of strong emission-line diagnostics in star-forming galaxies\altaffilmark{1}}
\author{
L.~L.~Cowie,$\!$\altaffilmark{2}
A.~J.~Barger,$\!$\altaffilmark{2,3,4}
A.~Songaila$\!$\altaffilmark{2}
}

\altaffiltext{1}{Based in part on data obtained at the W.~M.~Keck
Observatory, which is operated as a scientific partnership among
the California Institute of Technology, the University of
California, and NASA and was made possible by the generous financial
support of the W.~M.~Keck Foundation.}
\altaffiltext{2}{Institute for Astronomy, University of Hawaii,
2680 Woodlawn Drive, Honolulu, HI 96822.}
\altaffiltext{3}{Department of Astronomy, University of
Wisconsin-Madison, 475 North Charter Street, Madison, WI 53706.}
\altaffiltext{4}{Department of Physics and Astronomy,
University of Hawaii, 2505 Correa Road, Honolulu, HI 96822.}

\slugcomment{To be published in The Astrophysical Journal}

\begin{abstract}
We examine the redshift evolution of standard
strong emission-line diagnostics for H$\beta$-selected star-forming galaxies
using the local SDSS sample and a new $z=0.2-2.3$ sample
obtained from {\em HST\/} WFC3 grism and Keck DEIMOS and MOSFIRE data.
We use the SDSS galaxies to show
that there is a systematic dependence of the strong emission-line
properties on Balmer-line luminosity, which we interpret
as  showing that both the N/O abundance and the ionization parameter
increase with increasing line luminosity.  Allowing for
the luminosity dependence tightens the diagnostic diagrams
and the metallicity calibrations. The combined SDSS and
high-redshift samples show that there is no redshift evolution in the line properties
once the luminosity correction is applied, i.e., all
galaxies with a given $L({\rm H}\beta$) have similar strong emission-line
distributions at all the observed redshifts. We argue that the best
metal diagnostic for the high-redshift galaxies may be a luminosity-adjusted
version of the [NII]6584/H$\alpha$ metallicity relation.
\end{abstract}


\section{Introduction}
\label{secintro}

Over the past few years, there has been considerable
interest in determining the redshift evolution of 
rest-frame optical emission-line ratios in star-forming galaxies.
Emission-line properties can provide insight into the redshift 
evolution of stellar populations, gas densities, and metallicities
that cannot be obtained from spectral energy distributions alone.
The advent of multi-object near-infrared (NIR) spectrographs
has advanced the field considerably by making
it possible to obtain rest-frame optical spectra for substantial
samples of high-redshift ($z\sim1-3$) galaxies, which can then 
be compared with local samples
\citep[e.g.,][]{masters14,steidel14,wuyts14,shapley15,kewley15,yabe15}.

However, one has to be cautious with such comparisons, because
high-redshift galaxies have very different properties from the 
general population of local galaxies. Even local galaxies with 
masses similar to high-redshift galaxies may be quite dissimilar in their other 
properties, such as their star formation rates (SFRs), morphologies, 
and dust content \citep[e.g.,][]{cowie96}.

Similar concerns apply to the use of locally-calibrated strong-line metallicity 
diagnostics at high redshifts. Such calibrations are based on a variety of local 
galaxies whose properties are generally quite different from those of high-redshift galaxies.
Thus, these calibrations should not be applied uncritically at high redshifts. 

It is therefore clearly important to identify appropriate lower-redshift analogs for
any high-redshift galaxy comparisons. 
Moreover, if true analogs can be identified, then they 
could potentially be studied more easily. However 
they would have to be thoroughly tested against the properties of high-redshift 
galaxy samples to make sure they were representative.

\begin{inlinefigure}
\hspace*{-0.8cm}\includegraphics[angle=0,width=4in]{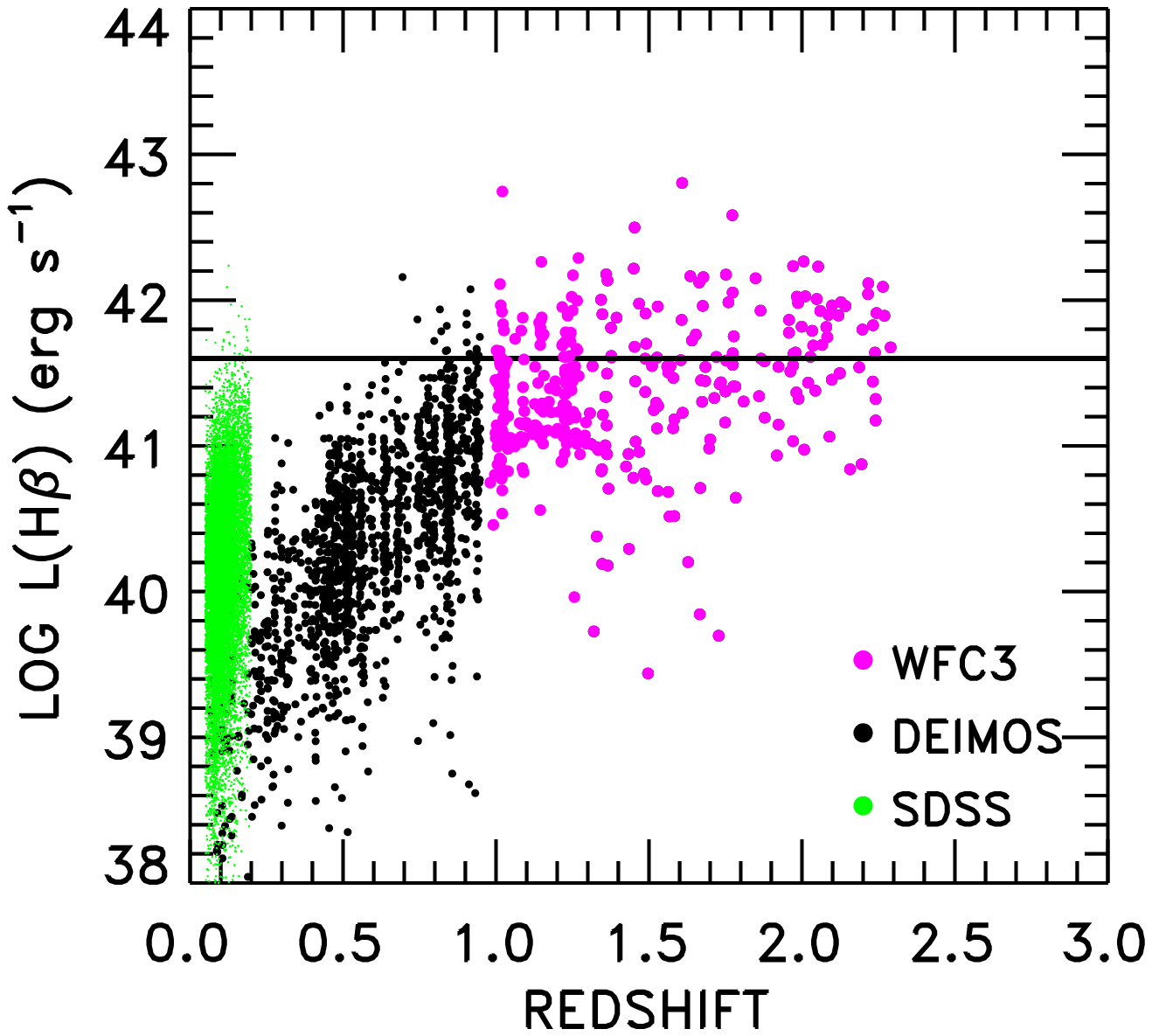}
\caption{Redshift evolution of the logarithm of H$\beta$ luminosity 
(green symbols---every 10th data point for SDSS galaxies with $z=0.05-0.2$; 
black circles---GOODS-N galaxies from DEIMOS observations; 
purple circles---GOODS-N galaxies from {\em HST\/} WFC3
grism observations). We have excluded a small number of 
galaxies in the GOODS-N whose X-ray luminosities identify them
as AGNs. The black line shows the median 
$\log L({\rm H}\beta) = 41.6$~erg~s$^{-1}$ for the $z>1.35$ galaxies. 
\label{sfr_z}
}
\addtolength{\baselineskip}{10pt}
\end{inlinefigure}

The most obvious difference between local and high-redshift galaxies 
is line luminosity (or, equivalently, for the Balmer lines, SFR). 
In Figure~\ref{sfr_z}, we
show the evolution of the H$\beta$ luminosity, $L$(H$\beta$), for galaxies
in the GOODS-N field (black and purple symbols) as compared
with SDSS galaxies with $z=0.05-0.2$  (green symbols).
(We describe both data sets in Section~\ref{secdata}.)
The most striking feature of this type of diagram is the well-known rapid 
rise in the line luminosity. 
For the H$\beta$ line, this rise reaches a plateau beyond $z\sim0.8$ 
with a maximum luminosity of $\log L{\rm (H}\beta) \sim 42.4$~erg~s$^{-1}$. 
A similar result has been noted for the Ly$\alpha$ line \citep{wold14}.

The rapid rise and plateau results --- combined with the lower luminosity 
bound set by the line-flux limits for $z\gtrsim0.8$ --- give a very high 
luminosity selection. 
For the GOODS-N sample, the $z>1.35$ population has a median
$\log L{\rm (H}\beta) = 41.6$~erg~s$^{-1}$ (black line). 
In contrast, Figure~\ref{sfr_z} shows that at low redshifts,
galaxies with such a high luminosity are extremely rare.

These changes may have profound implications for the interpretation of  
the commonly-used strong emission-line diagnostics, especially when we are comparing low-redshift and high-redshift galaxy samples.   
The BPT diagram, named after \citet*[][]{bpt81} \citep[see also][]{vo87},
plots $\log$(\bpty) vs.\ $\log$(\bptx) (hereafter, O3Hb versus N2Ha).
Originally developed to determine the ionizing mechanism in emission-line galaxies
(i.e., star formation  vs.\ active galactic nucleus activity),
it has recently been used to compare the excitation properties of high-redshift
star-forming galaxies with those of local star-forming galaxies. 
Unfortunately, however,
such comparisons are complicated by the fact that the BPT diagram is biased:
galaxies with a higher SFR (or, equivalently, Balmer line luminosity) lie above 
the average BPT locus at a given mass \citep{brinchmann08,salim14,newman14,
salim15}. 

\citet{juneau14} emphasized the importance of this effect in
interpreting the line properties of high-redshift galaxies.
They postulated that much of the apparent redshift evolution of the line properties
could in fact be a luminosity selection effect, in 
which only the high-SFR portion of the BPT star-forming galaxy locus at high 
redshifts (which is the only portion being sampled) is being compared with the median 
of the BPT star-forming galaxy locus locally. However, they
stressed that the rarity of the high line-luminosity
galaxies at low redshift means that they could be substantially
different from the high-redshift galaxies of the same luminosity,
making it hard to differentiate between selection bias and evolution.

However, as we shall show in the present paper, all
of the line ratios have  a strong and
well-defined dependence on the Balmer-line luminosity, 
suggesting that the low-redshift high-luminosity galaxies
may in fact be close analogs of the high-redshift galaxies.
Correcting for this luminosity dependence
considerably tightens all of the various diagnostic diagrams, and including
L(H$\beta$) as an additional parameter improves the metallicity estimators.
This correction also removes any redshift evolution in the line ratios
since the higher-redshift objects closely match lower-redshift samples
in the same Balmer-line luminosity range. 

Throughout, we use the observed $L$(H$\beta$),
without an extinction correction, as our variable, which allows for simple
corrections to any data set. We use $L$(H$\beta$)
rather than $L$(H$\alpha$), which is contaminated by [NII] in
low-resolution spectra.  However, since the
extinction correction smoothly varies with luminosity, we would expect
to see the same type of line-ratio dependence
on the extinction-corrected values or SFRs.
In the low-redshift SDSS data described in Section~\ref{sdssdata},
the mean extinction corrections
are almost independent of Balmer-line luminosity, and the results
of the present paper can be converted to $L$(H$\alpha$) using the
relation $\log L$(H$\alpha$) = $\log L$(H$\beta$) + 0.54.
The SFR for a \citet{kr2001}
IMF is given by a mean relation of $\log$ SFR
$= \log L$(H$\beta$) $-$ 40.50, where the SFR is in M$_\sun$~yr$^{-1}$
and $L$(H$\beta$) is in erg~s$^{-1}$ based on the SFRs of
\citet{brinchmann04}, as updated in the current data release.
(For a modified Salpeter IMF \citep{kenn98},
the SFR would be a factor of 1.5 higher.)
The intermediate-redshift data
also show an extinction that is independent of line
luminosity and gives a broadly similar correction to
$\log L$(H$\alpha$) = $\log L$(H$\beta$) + 0.58. The situation at
the highest redshifts ($z\sim2$) is less clear, and there
may be a dependence on line luminosity \citep{reddy15},
though this effect may be, at least partly, selection
bias. However, the average conversion from
H$\beta$ to H$\alpha$ is similar, with
$\log L$(H$\alpha$) $= \log L$(H$\beta$) + 0.54
for the galaxies with detected H$\beta$.
Some of the possible complexities in interpreting
the line luminosities are described in \citet{stasinska15b}.

In Section~\ref{secdata}, we present the SDSS data that we use and 
our highly complete optical and NIR spectroscopic sample of galaxies 
with $z=0-2.3$ in the GOODS-N field.
In Section~\ref{secsdss}, we focus on the BPT diagram from
the SDSS data and show that there is a simple dependence of
the star-forming galaxy locus on Balmer-line luminosity. 
In Section~\ref{seccomp}, we show that the positions of high-redshift
galaxies in the BPT and other diagnostic diagrams
are similar to those of SDSS
galaxies with comparable luminosities. This suggests
that we can use these latter objects as surrogates for the
high-redshift population. In Section~\ref{secdisc},
we use the properties of the high-luminosity SDSS galaxies to determine
what is causing the changes in the diagnostic diagrams with luminosity.
We summarize our results in Section~\ref{secsummary}.

\section{Data Sets }
\label{secdata}

\subsection{SDSS (z=0.05-0.2)}
\label{sdssdata}

Our local sample is taken from the Sloan Digital Sky 
Survey Data Release 7 \citep[SDSS DR7;][]{aba09}, restricted to
$z>0.05$ to minimize aperture effects. 
We use the  line fluxes from the Value Added Catalogs 
of the Max-Planck Institute for Astronomy (Garching) and
John Hopkins University 
(MPA/JHU)\footnote{http://www.mpa-garching.mpg.de/SDSS/DR7/}.
\citet{tremonti04} and Brinchmann et al. (2004) give details on the methods that 
were used to develop this catalog.  
Even in this redshift range we need to correct the
fluxes in the fibers to total fluxes in order to compare properly
with high-redshift samples. We follow the standard procedure
of multiplying the observed flux in the fiber by the ratio
of the total continuum flux to the continuum flux in the fiber.
This is clearly approximate since it assumes that the line emission
has the same spatial distribution as the continuum, but represents
our best estimate of the correction. The correction is near unity
for the galaxies with the highest line luminosities
though it can be significantly larger in
the low-luminosity objects, rising to an average correction of
approximately a factor of seven for a fiber $\log L$(H$\beta$)  in erg s$^{-1}$ of
39. Because the correction for the high-luminosity objects is small, the comparison
with the high-redshift high-luminosity objects is not significantly
affected by including this effect.
These spectra have been corrected for underlying stellar absorption.
We restrict consideration to sources with S/N $>10$ in H$\beta$ line flux 
and compute line luminosities and the various diagnostic 
line ratios, including O3Hb and N2Ha, for this sample.
We also consider only galaxies with $z=0.05-0.2$. We hereafter refer to
this as the ``full'' SDSS sample.

\subsection{DEIMOS (z=0.2-0.95)}
\label{deimosdata}

We made DEIMOS observations  in a number of 
runs between 2004 and 2015 with the goal of obtaining
consistently high-quality optical spectra of all of the 2850  $B<25$
galaxies in the uniformly-covered portion of the {\em HST\/}
GOODS-N field. We have now observed all but six of these galaxies,
and we have obtained robust redshifts for 2501 (88\%), many of
which are published in \citet{cowie04} and \citet{barger08}.
We used the 600~${\rm l\ mm}^{-1}$ grating,
giving a resolution of $3.5$~\AA\ and a wavelength coverage of
$5300$~\AA, which was also the configuration used in the KTRS 
observations \citep{wirth04}. We have incorporated archival data, particularly 
from KTRS, but many of these spectra have relatively
poor sky subtraction, and we re-observed them, whenever necessary,
to obtain higher-quality spectra. We centered the
spectra at an average wavelength
of $7200$~\AA, though the exact wavelength range for each
spectrum depends on the slit position. We broke each $\sim 1$~hr exposure
into three sub-exposures, stepping the galaxies along
the slit by $1\farcs5$ in each direction. We continuously
re-observed unidentified galaxies or galaxies with poor spectra, giving 
maximum exposure times of up to 7~hrs. The spectra were reduced  
following the procedures described in  \cite{cowie96}.
Our dithering procedure provides extremely high-precision sky subtraction, 
which is important for measuring accurate equivalent widths.

We measured the line widths, redshifts, amplitudes, and continuum level
for the suite of observable lines in the one-dimensional spectra using the 
MPFIT program of \citet{markwardt09}.
For the weaker lines, we measured only the amplitudes, and adopted the 
line-widths and redshifts from the strongest neighboring line (where available, 
[OIII]5007, H$\alpha$, and [OII]3727; for the latter, we measured both 
members of the doublet separately).

Because of the difficulties in making a precise spectrophotometric 
calibration of the multi-slit DEIMOS data, we used an indirect method
to calculate the fluxes. We matched the contributions of the 
lines plus continuum to the observed {\em HST\/} magnitude for
the corresponding broadband filter.  
We then determined errors for each line by fitting random positions
in the spectra in the neighborhood of the line.  Finally, 
we accounted for underlying absorption using fixed corrections of
3~\AA\ and 2~\AA\ for the equivalent widths 
of the H$\beta$ and H$\alpha$ lines, respectively. These match
the average corrections in the SDSS data.

We found 83 galaxies with H$\beta$ detected at $>10\sigma$ between 
$z=0.2-0.5$, where we can measure both O3Hb and N2Ha 
from the DEIMOS spectra. These provide
a nearly complete sample above $\log L($H$\beta$)=40.0~erg~s$^{-1}$
(see Figure~\ref{sfr_z}). We found 180 galaxies with H$\beta$ detected 
at $>10\sigma$ between $z=0.5-0.95$,
where only O3Hb can be measured from the DEIMOS spectra.
Here the completeness limit is $\log L($H$\beta$)=40.5~erg~s$^{-1}$. 
To get N2Ha determinations for these galaxies requires MOSFIRE spectra 
in the $Y$ and $J$ bands (see \S\ref{mosdata}).

\subsection{HST Grism ($z\ge$ 0.95)}
\label{grisdata}

For galaxies at $z\ge 0.95$, we measured the Balmer-line luminosities 
from the {\em HST\/} WFC3 G141 and G102 grism observations covering
nearly all of the GOODS-N field (PIs B.~Weiner and G.~Barro, respectively). 
While these data have now been incorporated into the
3D--HST sample \citep{brammer12}, we used our own extractions 
that we made from the raw data using the aXe software package supported 
by STScI. 
We extracted spectra for 5709 galaxies with AB magnitudes brighter 
than 24.5 in the F140W band. We formed one-dimensional spectra 
over the full galaxy areas, as defined by the corresponding
continuum image. 
We show a typical spectrum in Figure~\ref{sample_grism_spectrum}.

\begin{inlinefigure}
\hspace*{-0.6cm}\includegraphics[angle=0,width=3.4in]{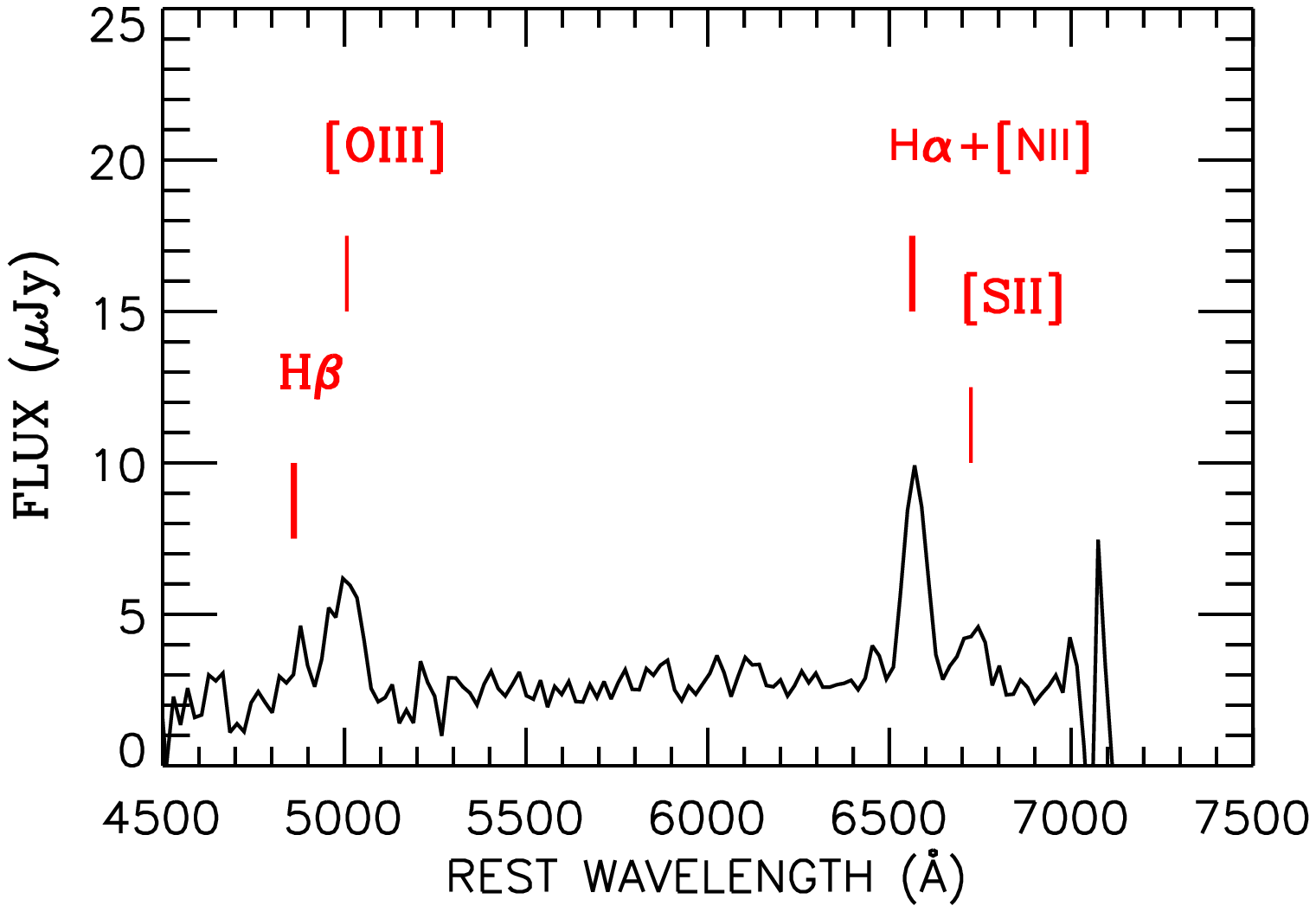}
\caption{WFC3 G141 grism spectrum of a $z=1.39$ galaxy with 
$\log L({\rm H}\alpha$+[NII])=42.16~erg~s$^{-1}$. 
The H$\alpha$ + [NII]6548,6584 blend is marked, as are the 
[OIII]4959,5007 + H$\beta$ complex and the [SII]6717,6734 blend.
\label{sample_grism_spectrum}
}
\addtolength{\baselineskip}{10pt}
\end{inlinefigure}

We can make precise line-flux measurements with the grism data,
since, in contrast with the ground-based data, there are no issues 
of slit losses or photometric conditions.
However, the resolution is low:  $R\sim130$ in the G141 grism 
over the wavelength range $1.1-1.7~\mu$m and $R\sim210$ in the
G102 grism over the wavelength range $0.8-1.15~\mu$m, with the
exact value depending on the size and shape of the galaxy.  This  
means that the H$\alpha$
line is blended with the [NII]6548,6584 lines,
and the [OII]3727 and [SII]6717,6734 doublets are 
unresolved (see Figure~\ref{sample_grism_spectrum}).
However, the lines in the [OIII]4959,5007 and H$\beta$ complex 
can be fitted to determine the H$\beta$ fluxes.  We did this in the following way:   
We measured the [OIII]5007 and H$\beta$ fluxes from the
one-dimensional spectra using MPFIT.  We then forced the [OIII]4959 line
to a value of 0.326 times the [OIII]5007 line and fitted all three lines
in the complex simultaneously using a single redshift and line-width.
For H$\alpha$, we measured the combined H$\alpha$+[NII] line flux. 
Once again, we corrected for underlying 
absorption using fixed corrections of
3~\AA\ and 2~\AA\ for the equivalent widths
of the H$\beta$ and H$\alpha$ lines.

We are interested in using the grism data primarily to find
galaxies in the redshift ranges
$z=1.38-1.74$ and $z=2.03-2.65$, where O3Hb and N2Ha
can be simultaneously measured from
ground-based MOSFIRE data (Section~\ref{mosdata}).
For redshifts $z\gtrsim1.52$, the H$\alpha$ line
is no longer observable in the grism data, and
we rely on H$\beta$, which is observable
in the G141 grism data through the whole lower redshift interval and 
out to $z=2.30$ in the upper interval.

Over $z=1.30-1.52$, both the H$\alpha$ and H$\beta$ lines are in the 
grism data, so we determine the completeness of the H$\beta$
selection by seeing at what point we start to miss H$\alpha$ sources in H$\beta$.
Based on this analysis, we adopt a flux selection
of $f$(H$\beta)=5\times10^{-18}$~erg~cm$^{-2}$~s$^{-1}$, which
corresponds to a luminosity limit of $\log L$(H$\beta$)=41~erg~s$^{-1}$
for $z=1.38-1.74$ and $\log L$(H$\beta$)=41.5~erg~s$^{-1}$
for $z=2.03-2.30$. These selections give samples of 43 and 28 galaxies,
respectively.

\subsection{MOSFIRE}
\label{mosdata}

In order to measure the standard diagnostic line ratios, 
we need higher-resolution data than the {\em HST\/} grism spectra
can provide. We therefore made follow-up observations
of the luminosity-selected sample using the MOSFIRE
NIR spectrograph on the Keck~I telescope \citep{mclean12}.
The ground-based NIR windows (and the corresponding MOSFIRE
filters) limit the redshift intervals
in which the line ratios can be measured. We illustrate this
in Figure~\ref{zmos_kmag}, where for each of the N2Ha
and O3Hb ratios, we show the redshift interval where the ratio
can be measured (black for optical from DEIMOS, gold
for $Y$-band, blue for $J$-band, green for $H$-band, 
and purple for $K$-band). The gray shading shows where
it is possible to measure both ratios via some combination of
instruments and bandpasses;
this encompasses nearly the full $z=0-1$ interval, as well as
the intervals $z=1.38-1.74$ and $z=2.03-2.65$.

We obtained MOSFIRE observations of the H$\beta$ luminosity-selected 
galaxies during three two-night runs from 2013 to 2015.
We focused on galaxies with redshifts  that allowed 
measurement of all of the line diagnostics.
However, MOSFIRE can configure up to 46 slits and, since the target
sources were too sparse to fill the masks, we also
observed high-luminosity galaxies outside these redshifts,
together with filler targets. We used a $0\farcs7$
slit width and made observations in the $J$ (24~min exposure),
$H$ (16~min), and $K$ (24 min) bands with an
ABBA stepping pattern, where the A position is $1.25''$ below the nominal
position, and the B position is $1.25''$ above the nominal position. 
We carried galaxies into subsequent masks when additional
exposure time was required. In addition, we searched
the KOA archive for all publicly-available data
(obtained prior to May 2014)
on the GOODS-N. In total, our own observations plus the archival
data provided spectra for
404 galaxies, of which 177 have identifiable NIR emission
lines that make it possible to measure redshifts. We
show the galaxies with measured MOSFIRE redshifts
in Figure~\ref{zmos_kmag} (red squares).


\begin{inlinefigure}
\hspace*{-0.2cm}\includegraphics[angle=0,width=3.8in]{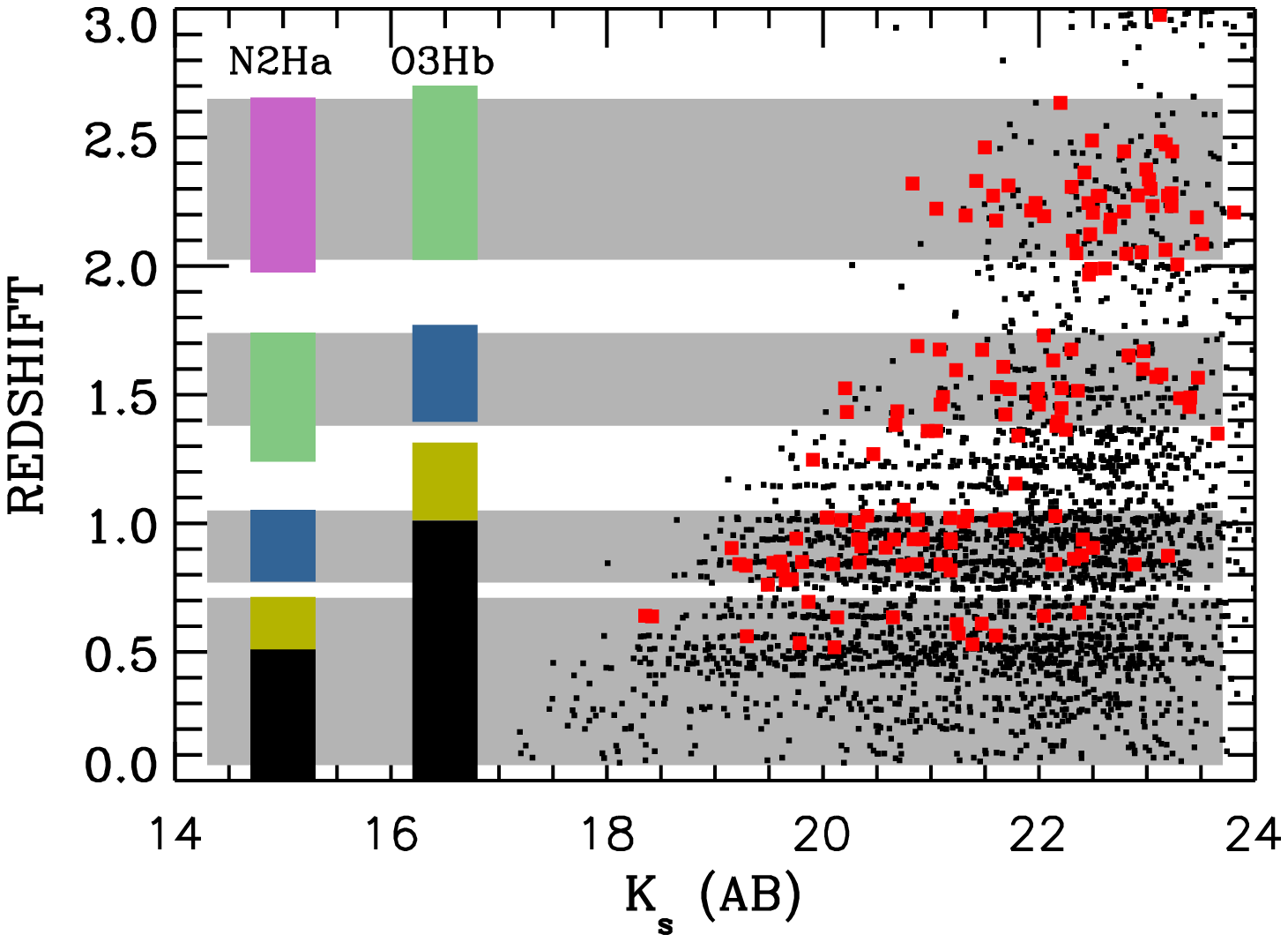}
\vskip -0.5cm
\caption{Redshift  vs.\ $K_s$ magnitude for the GOODS-N (black squares). 
Galaxies with MOSFIRE spectra from which the
redshift can be measured are shown with larger red squares.
The redshifts at which the N2Ha and O3Hb ratios (colored bars) 
can be measured are shown on the left. The colors mark the MOSFIRE
filter in which the lines lie (gold---$Y$, blue---$J$, green---$H$, and
purple---$K$), while black marks where the lines are covered by the
DEIMOS spectra. The gray shading shows the redshift ranges in which 
some combination of instruments and bandpasses 
allows both ratios to be measured.
(We ignore the tiny region around $z\sim 1.2-1.3$ where the green and gold 
bars overlap.)
\label{zmos_kmag}
}
\addtolength{\baselineskip}{10pt}
\end{inlinefigure}

\vskip -0.5cm

\begin{inlinefigure}
\hspace*{-1.9cm}
\includegraphics[angle=0,width=4.2in]{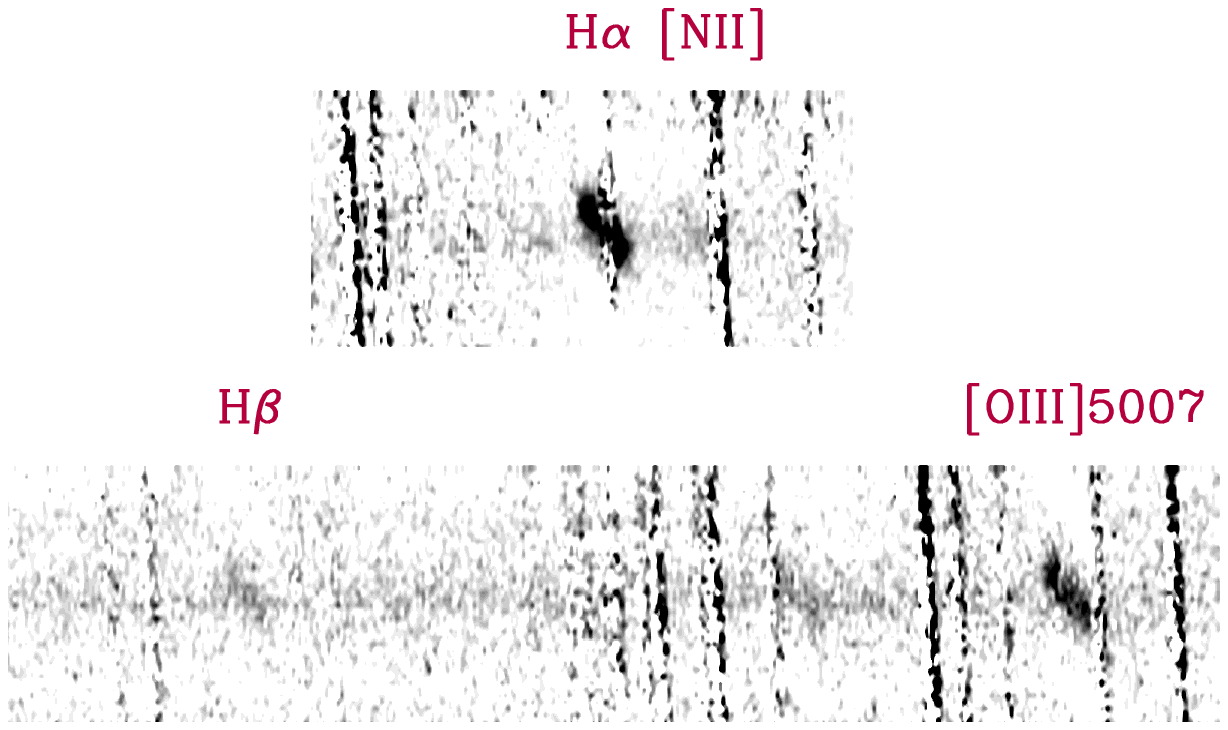}
\vskip -1cm
\caption{Spectral images around the H$\alpha$ + [NII] complex (upper
spectrum) and around the H$\beta$ + [OIII] complex (lower spectrum) for a galaxy
at $z=1.432$ with $\log L(H\beta$)=42.5~erg~s$^{-1}$. Note the shape of the
lines in the wavelength ($x$-axis)  vs.\ spatial position on the slit ($y$-axis),
and also the contamination of portions of the spectral features by strong night 
sky lines.
\label{mosfire_spectral_image}
}
\addtolength{\baselineskip}{10pt}
\end{inlinefigure}

We extracted the spectra using our own IDL-based package.
Each exposure was flat-fielded using dome flats taken in the afternoon,
and cosmic rays and bad pixels were flagged out.
We then adjusted the normalization of each exposure in the ABBA 
pattern to provide an optimal matching of the strength
of the night sky lines in each of the four exposures.
Both the A exposures and the B exposures were summed.
We then subtracted the summed B image from the summed A image
to form a differenced image, which removes the sky lines.
We shifted the differenced image
by $2\farcs5$ so that the positive A in the shifted differenced
image lay at the position of the negative B 
in the unshifted differenced image.
We then subtracted the unshifted differenced image from the shifted 
differenced image to form the final ABBA image, which contains the 
positive spectrum together with two negative residuals on either
side (Figure~\ref{mosfire_spectral_image}).

We also formed the corresponding two-dimensional
sky image by summing the ABBA frames shifted in the same way.
We made fits to the sky lines together with calibration
spectra taken in the the afternoon to form 
the wavelength calibration as a function of spatial
position in the two-dimensional image, and then we combined the multiple
exposures in each bandpass.
Because of the extreme stability of MOSFIRE, we
did not find it necessary to further register
these images in either the spatial or wavelength axes. 
We show sections of a typical two-dimensional image
in Figure~\ref{mosfire_spectral_image}.

In contrast to the DEIMOS spectra, the continua in the MOSFIRE 
spectra are often too weak to measure equivalent widths, so we
measured flux ratios directly. 
Since the flux ratios are only of neighboring lines, 
we do not need accurate spectrophotometry. To account for
underlying absorption, we measured equivalent widths in the grism 
data and applied fixed corrections of 3~\AA\ and 2~\AA\ for H$\beta$ 
and H$\alpha$, respectively.
Where H$\alpha$ was not covered by the grism data, we computed 
the H$\alpha$ equivalent width from H$\beta$ using 
the median value of the EW(H$\alpha$)/EW(H$\beta$) ratio determined from 
the sources where both were measured. Most of the galaxies have
corrections which are close to the median values of $0.89$ in
[OIII]/H$\beta$
 and $0.98$ in [NII]/H$\alpha$. Thus the use of a single correction
value for each line ratio as in \cite{steidel14} 
(who used 0.85 for [OIII]/H$\beta$
 and 1 for [NII]/H$\alpha$) provides a good approximation.
Removing the absorption correction  would raise O3Hb
by 0.07 in the high-redshift galaxies.

\begin{figure*}[tbh]
\includegraphics[angle=0,width=7.8in]{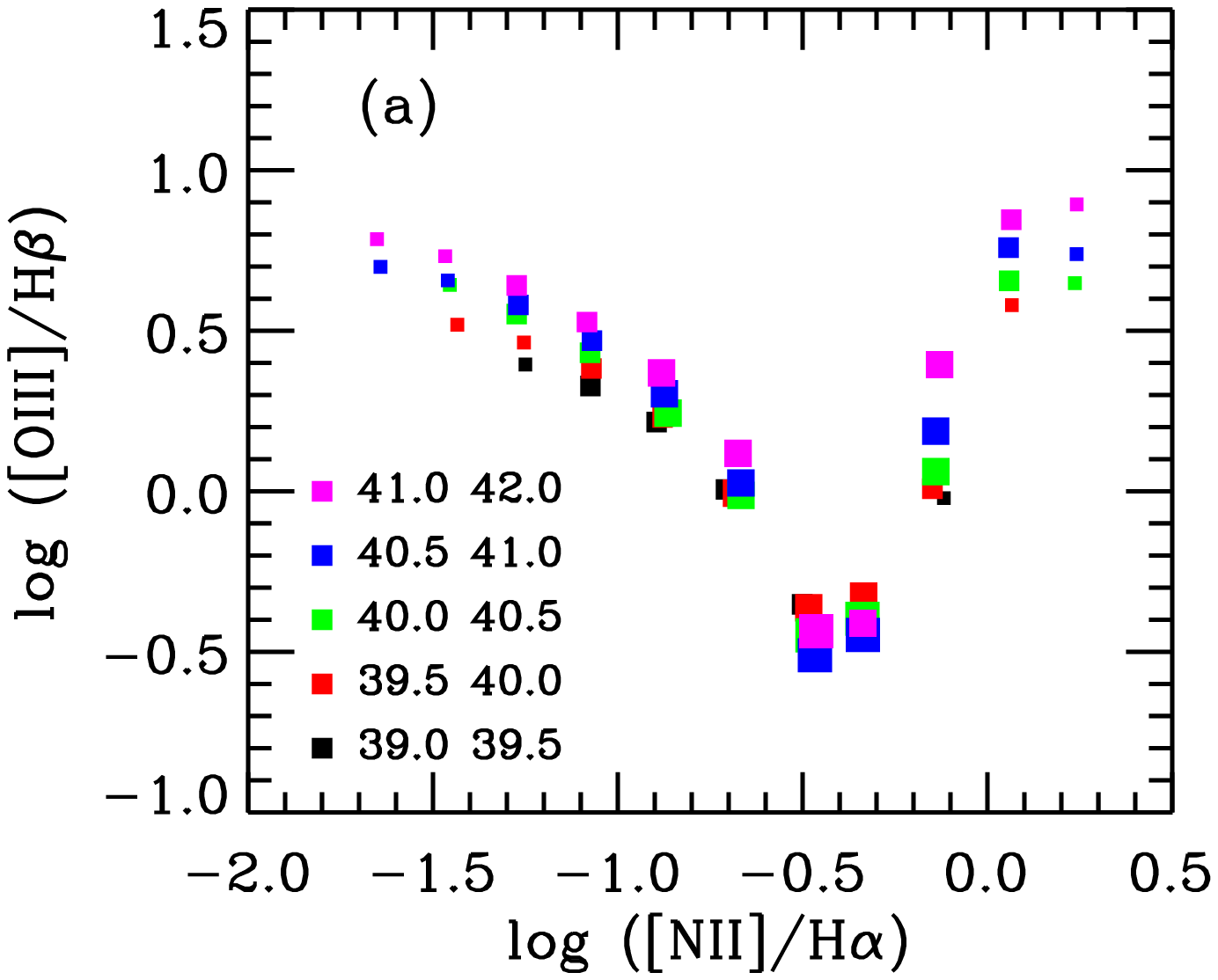}
\vskip -0.2cm
\hspace*{1cm}\includegraphics[angle=0,width=3.7in]{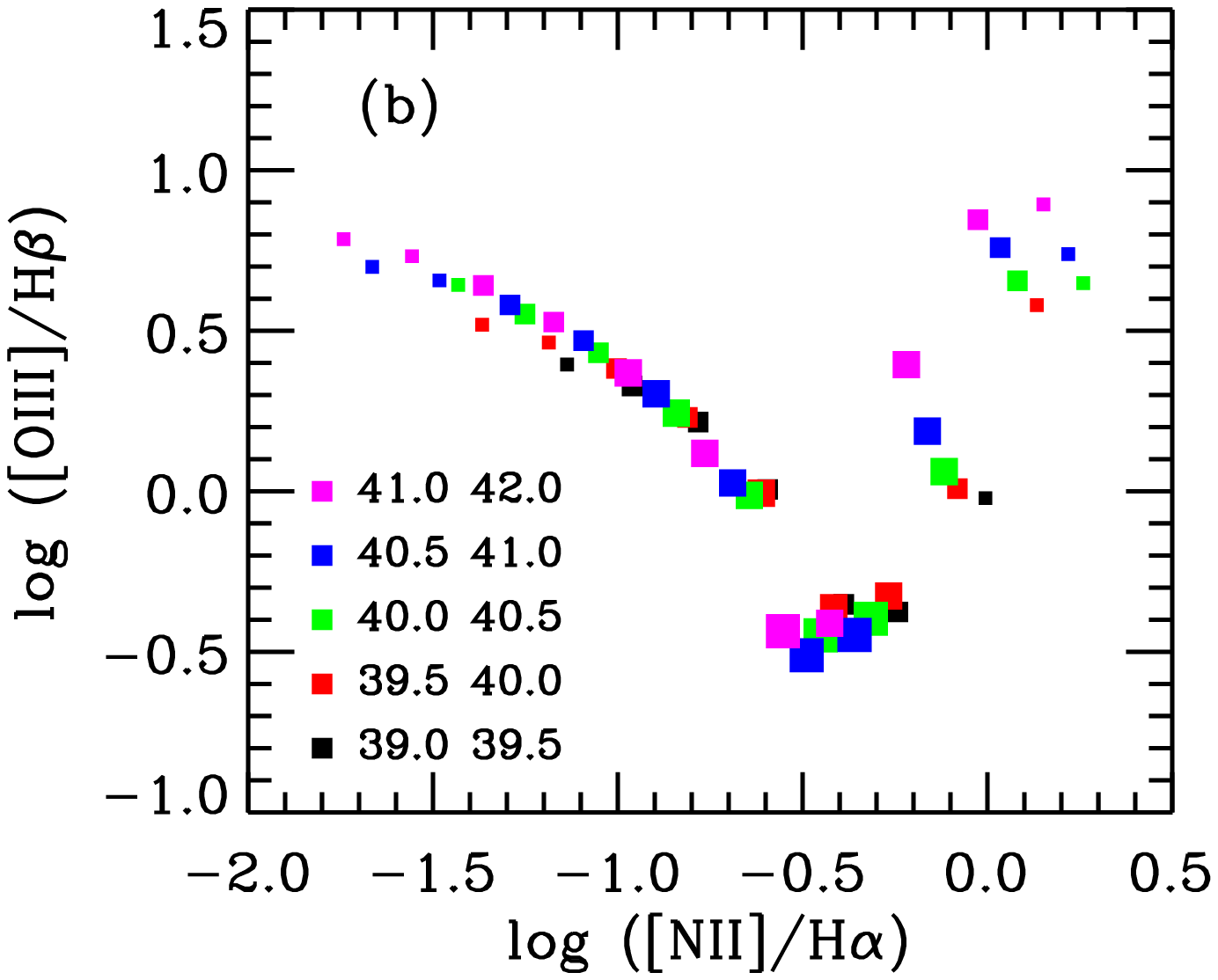}
\vskip -6.75cm
\hspace*{9cm}\includegraphics[angle=0,width=3.7in]{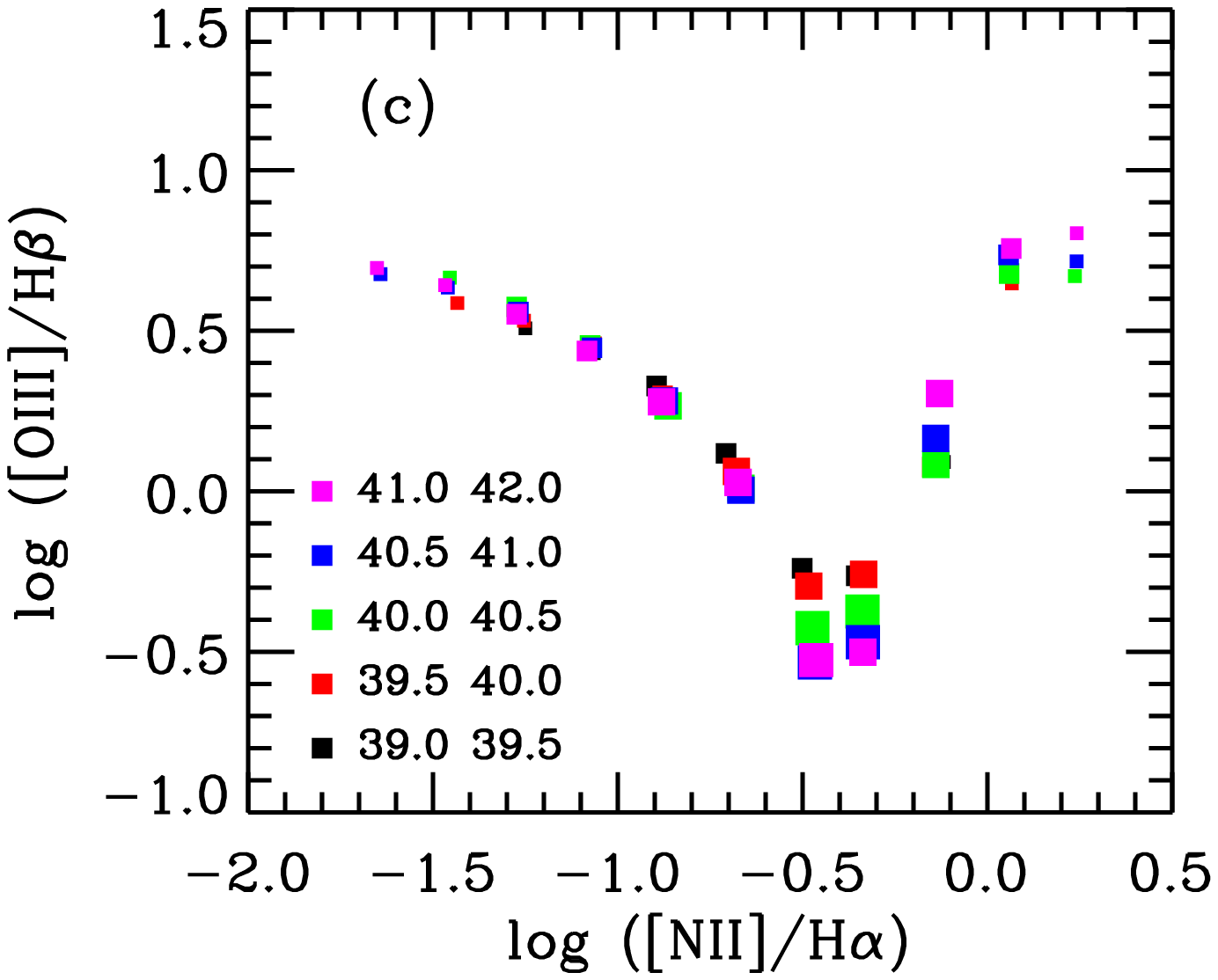}
\caption{(a) BPT diagram for the full SDSS sample 
in intervals of $\log L$(H$\beta$) (colors). 
We show the median values of 
O3Hb in 0.2~dex bins of N2Ha (squares). The sizes of the squares
represent the number of sources in each
bin, ranging from $10-100$ (smallest), $100-1000$, $1000-10000$,
to $10000-100000$ (largest). 
In the star-forming galaxy locus (the left wing of the BPT diagram; N2Ha~$<-0.5$),
the normalization of O3Hb relative to N2Ha rises with $L($H$\beta$). 
In (b) and (c), respectively, we show the effects of introducing an offset in N2Ha 
using Equations~\ref{locus1} and \ref{locus2} and in
O3Hb using Equation~\ref{locus3} to normalize to 
$\log L({\rm H}\beta)=40.5$~erg~s$^{-1}$.
Either offset approximately aligns the BPT star-forming galaxy locus.
\label{sdss_bpt_by_sfr}
}
\addtolength{\baselineskip}{10pt}
\end{figure*}

We also work with the two-dimensional spectral images rather 
than extracting one-dimensional spectra. There are two reasons for
following this course, which we illustrate
in Figure~\ref{mosfire_spectral_image}. 
The first is that the two-dimensional spectral images often
show resolved velocity structure. In these cases,
extracting a one-dimensional spectrum adds unnecessary
noise relative to working directly with the two-dimensional
spectral image. The second is the strength of the night-sky 
lines in the NIR, which can greatly increase the noise. 
In the upper panel of Figure~\ref{mosfire_spectral_image},
we see that portions of the H$\alpha$  and [NII]6584 lines are severely
contaminated in this way. By extracting the fluxes from
the two-dimensional spectral images, we can avoid these regions. 

For both N2Ha and O3Hb, we are measuring a weaker line
([NII]6584 or H$\beta$) relative to a stronger neighboring
line (H$\alpha$ or [OIII]5007).
We made an optimal extraction from the two-dimensional
spectral image by forming an integral of the two-dimensional
spectral image of the weaker line weighted by the two-dimensional 
intensity of the stronger line, excluding any regions covered in either 
by the strongest night-sky lines, and dividing
by the integral of the two-dimensional spectral image of the stronger
line with the same weighting. Once
again, we determined the noise levels by randomizing the wavelength positions
of the weaker line and measuring the ratio at these positions. 

\section{Local Relations from the SDSS Data}
\label{secsdss}

We start by re-examining the dependence of the BPT star-forming galaxy locus 
(the left wing of the BPT diagram) on
Balmer-line luminosity and redshift using the SDSS sample.
In Figure~\ref{sdss_bpt_by_sfr}(a), we show the BPT diagram for the full
SDSS sample in intervals of $\log L$(H$\beta$) ranging from
$39-39.5$~erg~s$^{-1}$ (black squares) to $41-42$~erg~s$^{-1}$ (purple squares). 
For each $\log L$(H$\beta$) interval, we show the median values of O3Hb in
0.2~dex bins of N2Ha. The sizes of the squares represent the number of sources in 
each bin. 

We immediately see the known result that at higher L(H$\beta$) luminosities,
O3Hb is high at a given N2Ha and N2Ha is high at a given O3Hb  
\citep{brinchmann08,juneau14,salim14}.
More profoundly, however, we see that the shape of the star-forming galaxy
locus remains roughly invariant
with $\log L$(H$\beta$).  It is only the normalization of O3Hb relative to N2Ha 
that is changing as a function of $\log L$(H$\beta$).

Motivated by Figure~\ref{sdss_bpt_by_sfr}(a), we
fitted the dependence of O3Hb on N2Ha and $L({\rm H}\beta$) 
for the star-forming galaxy locus (N2Ha~$<-0.5$) using
the MPFIT2DFUN two-dimensional fitting routine from \citet{markwardt09}.
We adopted a third-order fit for the dependence on N2Ha and
a linear fit for the dependence on $\log L({\rm H}\beta$). Including the
luminosity dependence by shifting N2Ha, we find
\begin{equation}
x={\rm N2Ha} -0.09(\log L({\rm H}\beta) -40.5) \,,
\label{locus1}
\end{equation}
and
\begin{equation}
{\rm O3Hb} = -1.85-4.02x -2.18 x^{2} \\
 -0.42 x^{3}  \,.
\label{locus2}
\end{equation}
Alternatively, we can include the luminosity dependence by shifting O3Hb, 
giving
\begin{eqnarray}
\nonumber
{\rm O3Hb} = -1.85-4.02({\rm N2Ha}) -2.18({\rm N2Ha})^{2}\\
-0.42({\rm N2Ha})^{3} +0.09(\log L({\rm H}\beta) -40.5)  \,.
\label{locus3}
\end{eqnarray}

In Figures~\ref{sdss_bpt_by_sfr}(b) and (c), we show the effects of 
applying the above luminosity offsets in N2Ha and O3Hb, respectively,  
to the SDSS data to normalize to $\log L({\rm H}\beta)=40.5$~erg~s$^{-1}$. 
Either offset aligns the star-forming galaxy loci across the luminosity range
and, based solely on the BPT diagram, variation in either variable
or a combination of both could produce the measured offsets.

We find that the formal errors on the coefficients are extremely small. However,
when we vary the fitting ranges  for both redshift and N2Ha, and when we
adjust the $L({\rm H}\beta$) S/N away from the adopted value
of 10 (see Section~\ref{sdssdata}), we find larger systematic errors.
In particular, the $\log L$(H$\beta$) coefficient in both sets of equations 
can range from 0.08 to 0.10, depending on these choices. 

The luminosity offset is not simply a shift in the medians, as 
we illustrate in Figure~\ref{sdss_bpt_highsfr_byz_points}, where
we plot the BPT diagram for the individual SDSS points in
$\log L({\rm H\beta})$ intervals of (a)
$41.25-41.75$ and (b) $40.25-40.75$~erg~s$^{-1}$,
each separated into two redshift intervals
(black for $z=0.1-0.2$ and red for $z=0.05-0.1$).

\begin{inlinefigure}
\includegraphics[angle=0,width=3.6in]{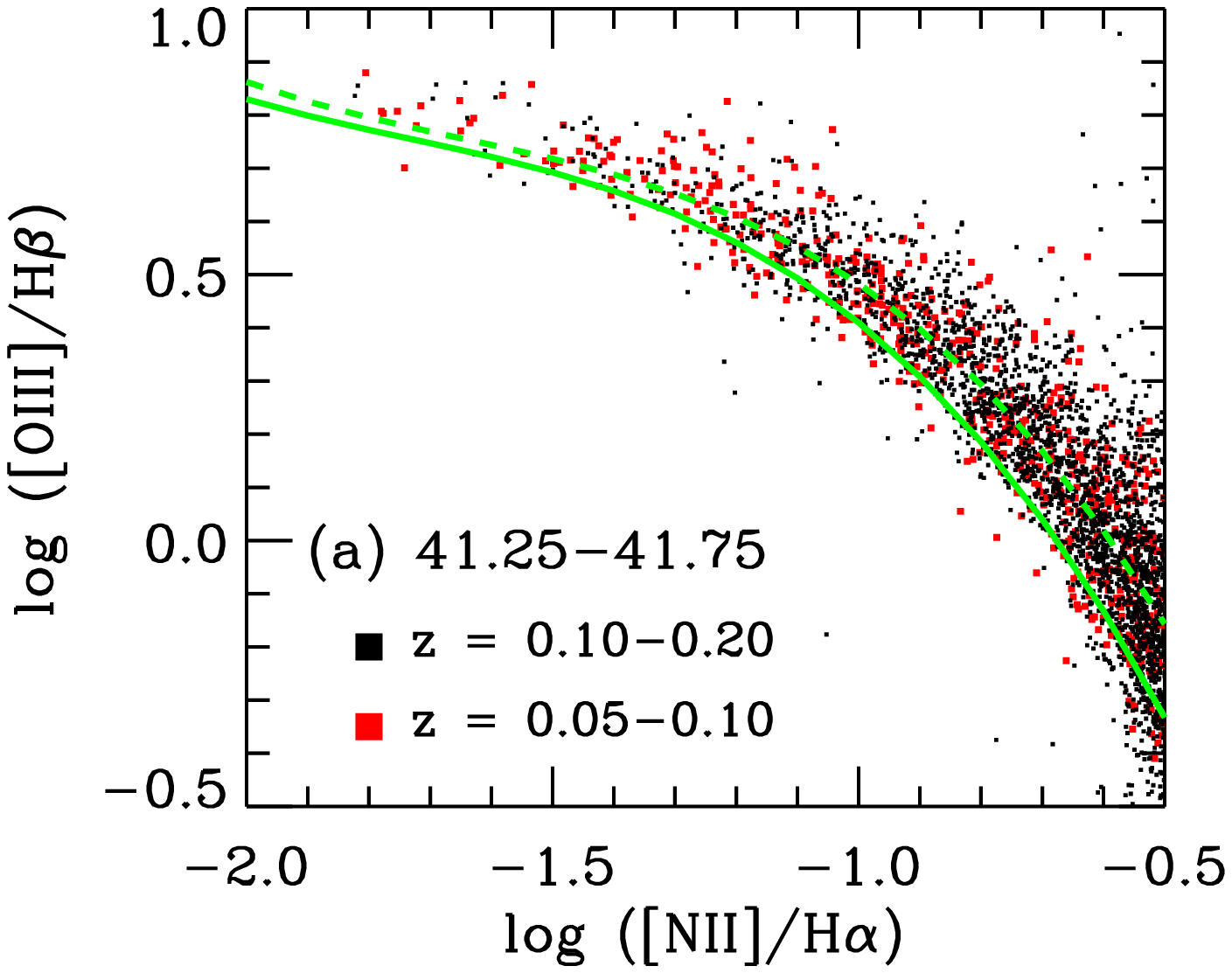}
\includegraphics[angle=0,width=3.6in]{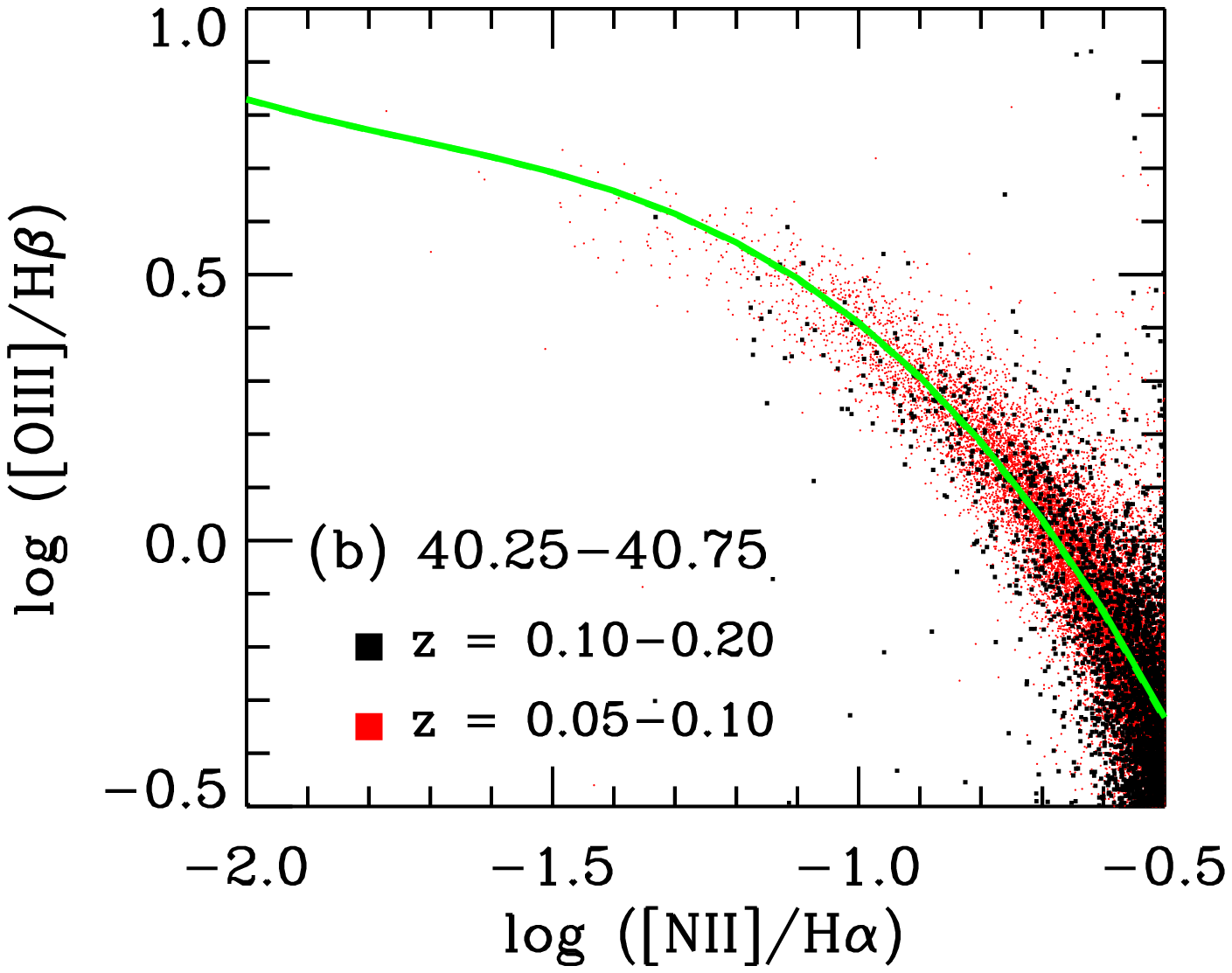}
\includegraphics[angle=0,width=3.6in]{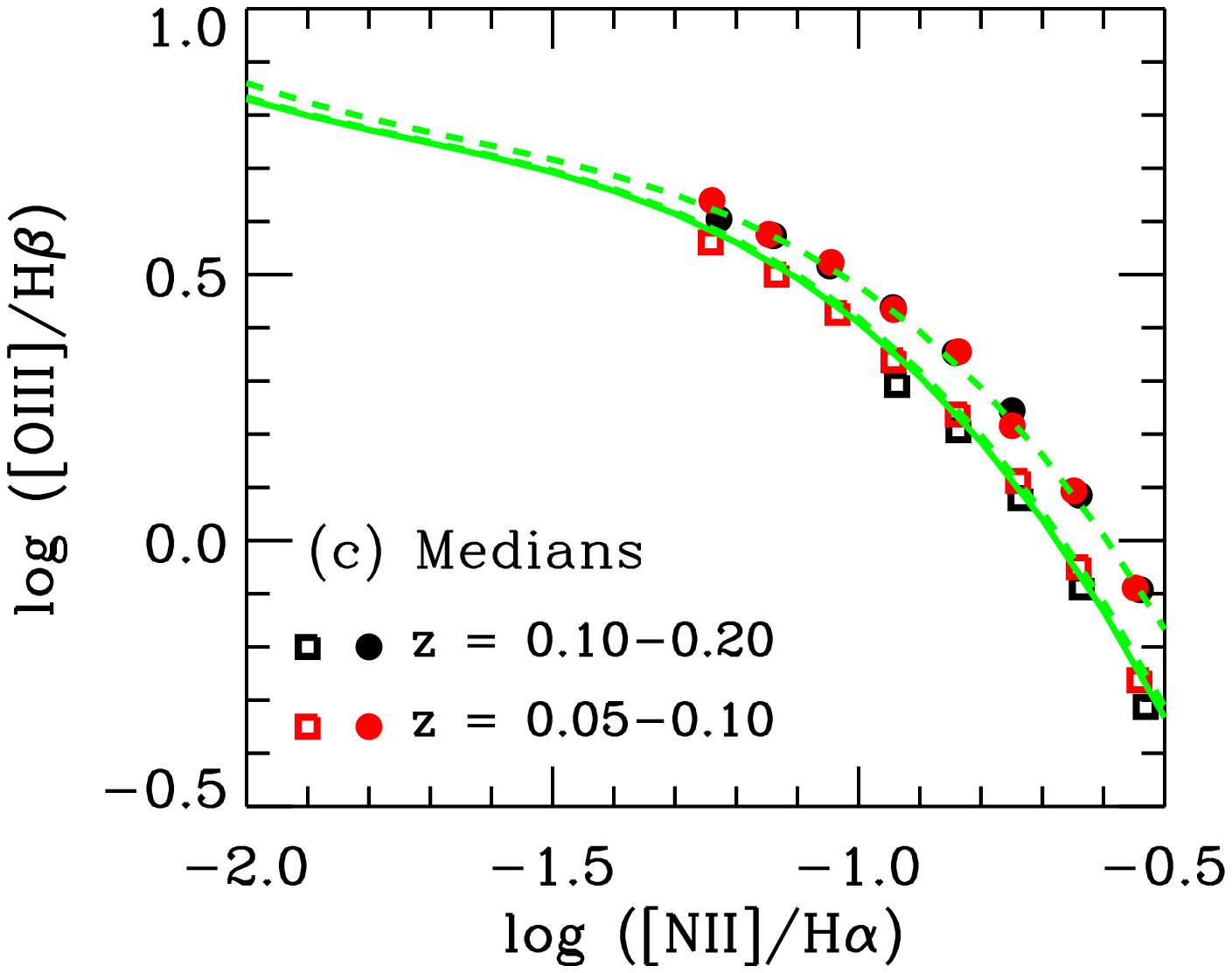}
\caption{BPT diagrams for the full SDSS sample in two luminosity intervals
of $\log L({\rm H}\beta)$, (a) $41.25-41.75$
and (b) $40.25-40.75$~erg~s$^{-1}$, each separated
into two redshift intervals 
(red squares---$z=0.05-0.1$; black squares---$z=0.1-0.2$).
The green solid curve in both panels shows the fit to the star-forming 
galaxy locus of the full SDSS sample at the median 
$\log L({\rm H}\beta)=40.5$~erg~s$^{-1}$, while
the green dashed curve in panel a shows the luminosity-offset locus using
Equations~\ref{locus1} and \ref{locus2}, which vary N2Ha,
evaluated at $\log L({\rm H}\beta)$ of
41.5. Panel (c) shows the median values of N2Ha for the two samples with
the circles showing the high luminosity interval of panel (a) and the
open squares the lower luminosity interval of panel (b).
\label{sdss_bpt_highsfr_byz_points}
}
\addtolength{\baselineskip}{10pt}
\end{inlinefigure}

There is essentially no difference between the 
source distributions for the two redshift intervals in either panel.
There is, however, an increase in N2Ha relative to O3Hb for the
higher-luminosity points in (a). This can clearly be seen by 
plotting the fit to the star-forming galaxy locus of the full SDSS
sample at the median $\log L({\rm H}\beta)=40.5$~erg~s$^{-1}$
(green solid curve), along with the luminosity-offset loci (green dashed curve)
determined by evaluating Equations~\ref{locus1} and \ref{locus2},
which vary N2Ha, at $\log L({\rm H}\beta)$ of
(a) 41.5~erg~s$^{-1}$ and (b) 40.5~erg~s$^{-1}$.  As expected,
the higher-luminosity points in (a) lie uniformly above the green solid 
curve but generally along the green dashed curve, while
the lower-luminosity points in (b) are consistent with the green solid curve, 
which is nearly identical to the green dashed curve.
The agreement may be more clearly seen in panel (c) where we show
the medians in each  luminosity and redshift interval compared with the fits.

\section{Comparison with Higher-Redshift Samples }
\label{seccomp}

\begin{figure*}
\centerline{\includegraphics[angle=0,width=6.0in]{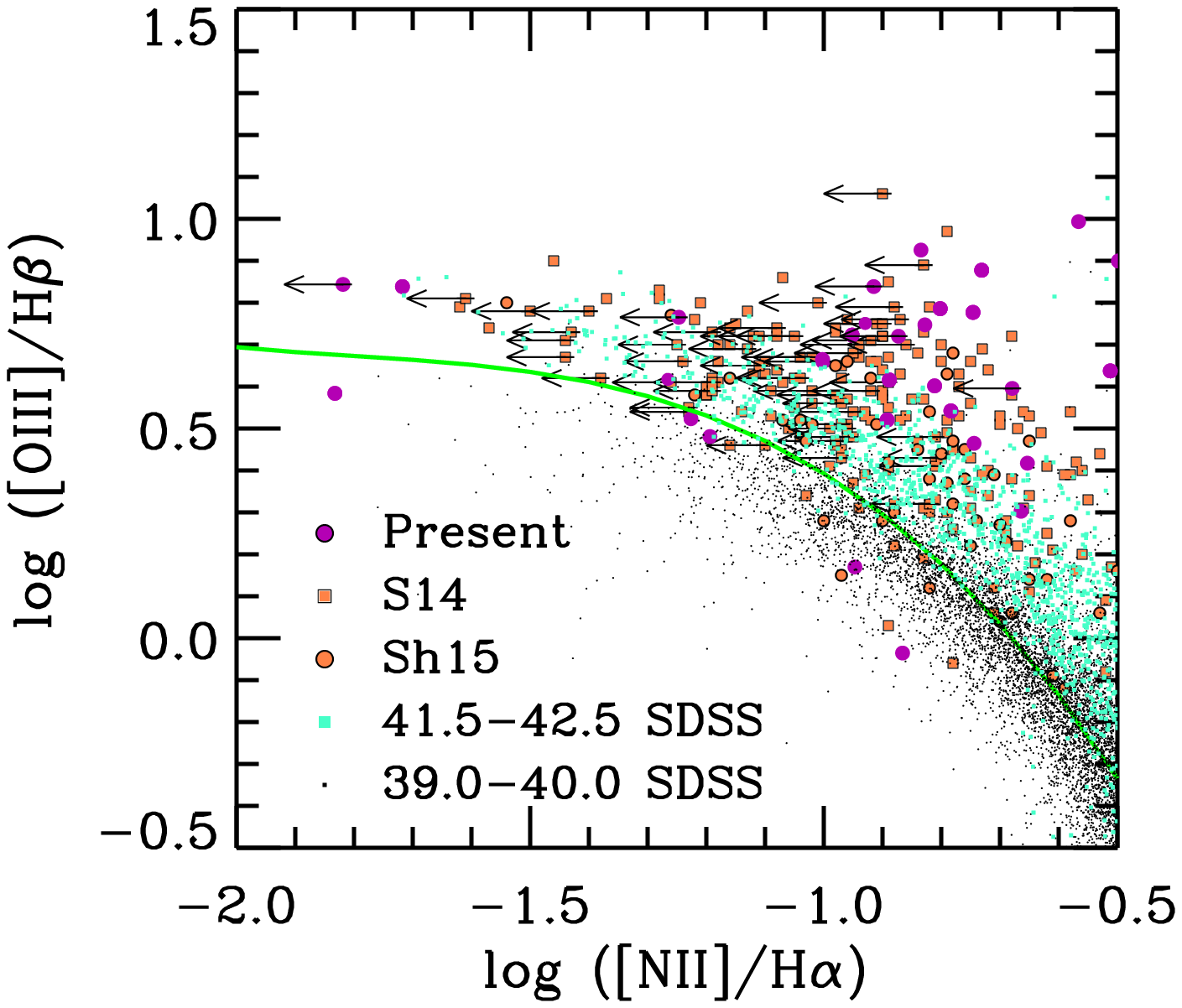}}
\caption{
BPT diagram for the full SDSS sample in two intervals of $\log L({\rm H}\beta)$ 
(black dots---$39.0-40.0$~erg~s$^{-1}$;
light green squares---$41.5-42.5$~erg~s$^{-1}$).
The green solid curve shows the fit to the star-forming
galaxy locus (N2Ha~$<-0.5$) of the full SDSS sample at the median
$\log L({\rm H}\beta)=40.5$~erg~s$^{-1}$.
For comparison, we show our data (every source with
O3Hb $>3\sigma$; purple circles), along with the high-redshift data from
\citet[orange squares]{steidel14}
and \citet[orange circles]{shapley15}. 
Sources with only upper limits on N2Ha are plotted at the 
$2\sigma$ values with leftward pointing arrows.
Some of the sources are likely AGNs, given their locations on the diagram.
For the star-forming galaxy locus,
the positions in the diagram of the high-redshift galaxies closely match those
of the higher-luminosity SDSS galaxies.
\label{all_bpt}
}
\end{figure*}

\begin{figure*}
\centerline{\includegraphics[angle=0,width=6.0in]{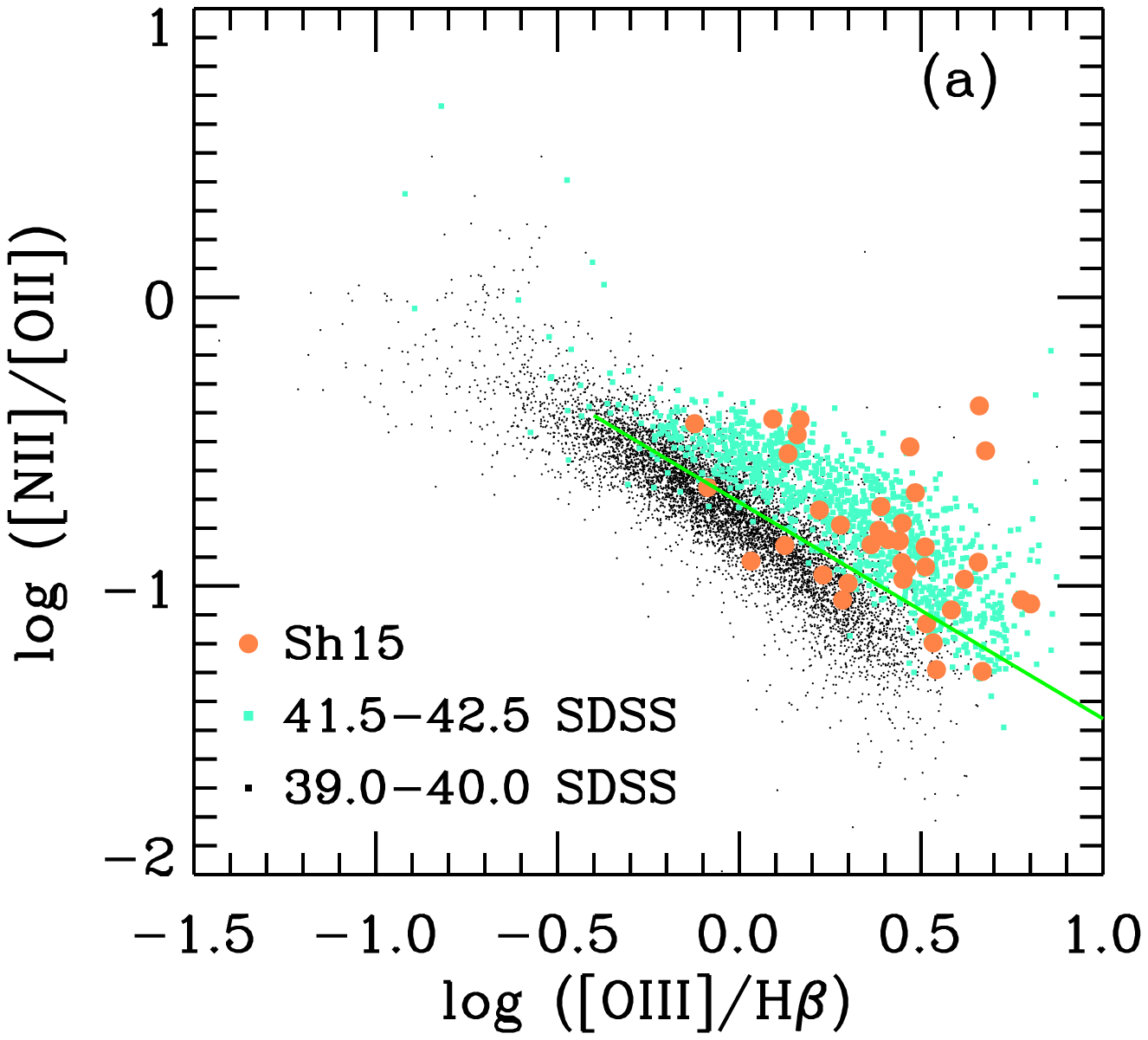}}
\centerline{\includegraphics[angle=0,width=6.0in]{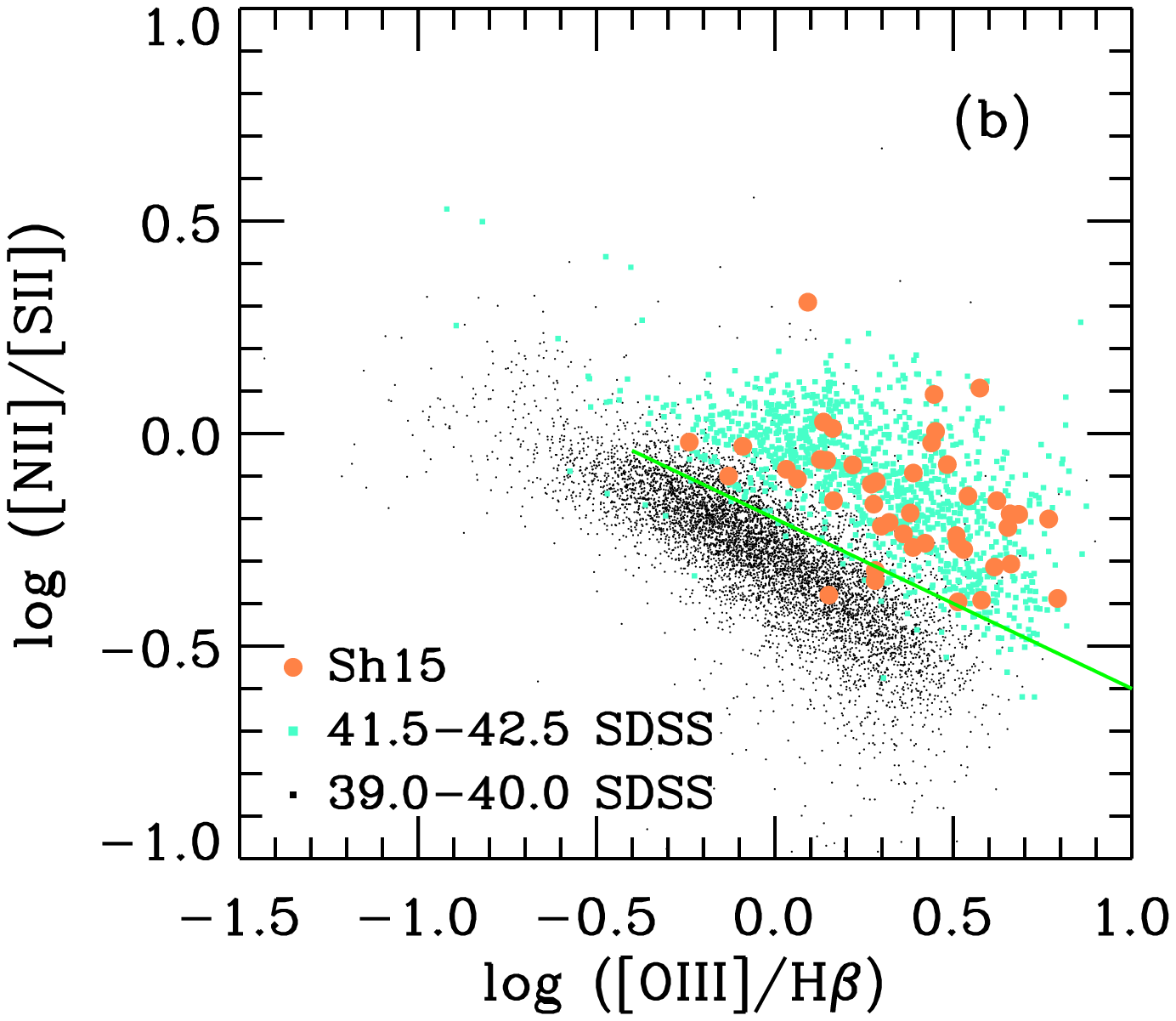}}
\caption{
(a) N2O2 and (b) N2S2  vs.\ O3Hb for the full SDSS sample on the
star-forming galaxy locus of the BPT diagram (N2Ha~$<-0.5$)
in two intervals of $\log L({\rm H}\beta)$ 
(black dots---$39.0-40.0$~erg~s$^{-1}$;
light green squares---$41.5-42.5$~erg~s$^{-1}$).
The green solid lines show linear fits to the star-forming
galaxy locus of the full SDSS sample at the median 
$\log L({\rm H}\beta)=40.5$~erg~s$^{-1}$.
The orange circles show the high-redshift galaxies from \cite{shapley15}, 
which closely match the positions in the diagrams of the 
higher-luminosity SDSS galaxies. The luminosities of the high
redshift galaxies lie between $\log L$(H$\beta$) = 41.1-42.6 with
85$\%$ lying in the 41.5-42.5 range (Reddy et al. 2015).
Extinction corrections are included in N2O2 for all of the samples, because 
of the large wavelength separation between the lines.
\label{alice_fig}
}
\end{figure*}

\begin{figure}
\hspace*{-0.5cm}\includegraphics[angle=0,width=4.3in]{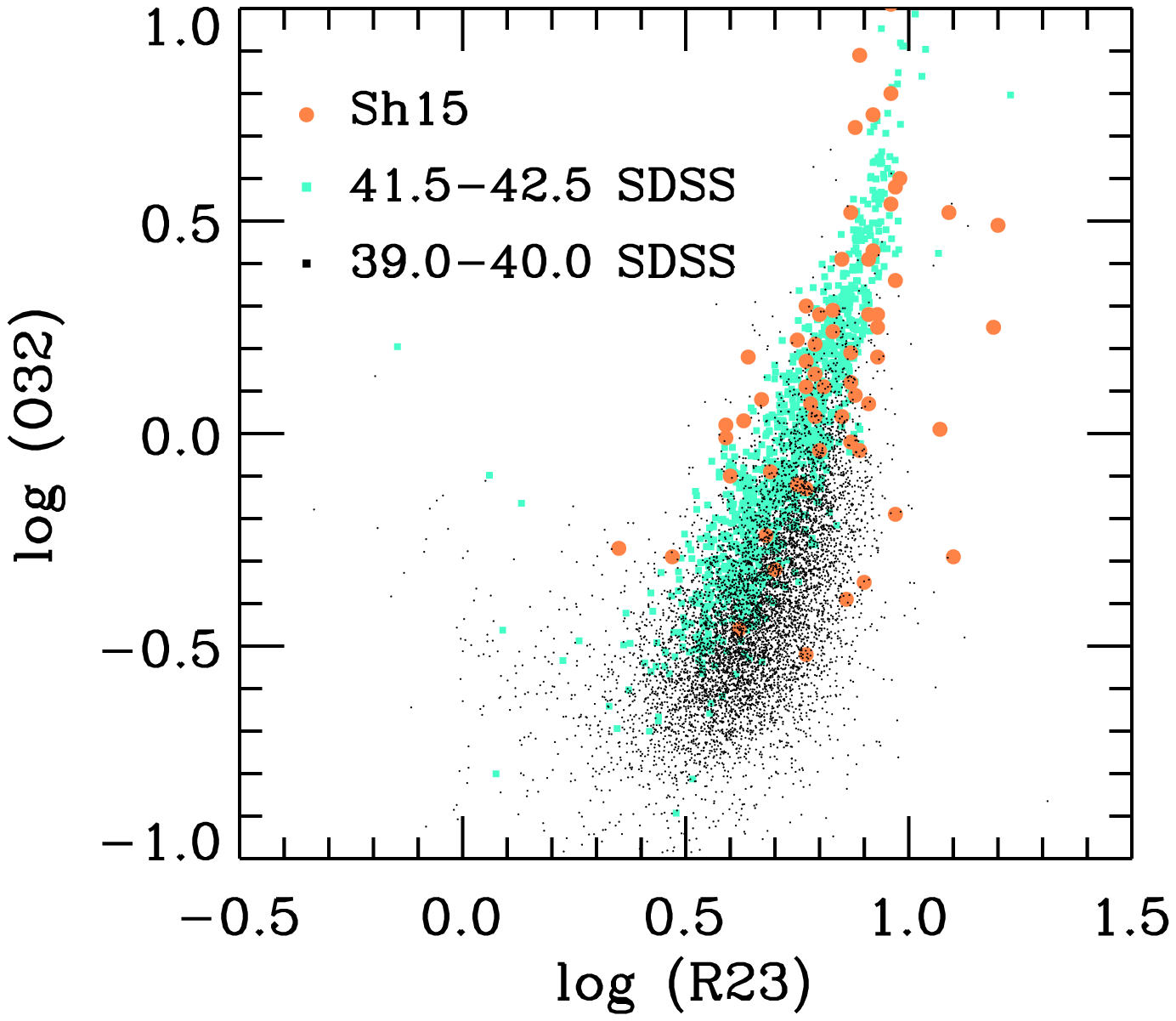}
\caption{
O32  vs.\ R23 for the full SDSS sample on the
star-forming galaxy locus of the BPT diagram (N2Ha~$<-0.5$)
in two intervals of $\log L({\rm H}\beta)$ 
(black dots---$39.0-40.0$~erg~s$^{-1}$;
light green squares---$40.5-41.5$~erg~s$^{-1}$).
The orange circles show the high-redshift galaxies from \cite{shapley15}, 
which closely match the positions in the diagram of the 
higher-luminosity SDSS galaxies.
Extinction corrections are included in both ratios for all of the samples.
\label{r23_o32}
}
\addtolength{\baselineskip}{10pt}
\end{figure}

\begin{figure*}
\centerline{
\includegraphics[angle=0,width=6.0in]{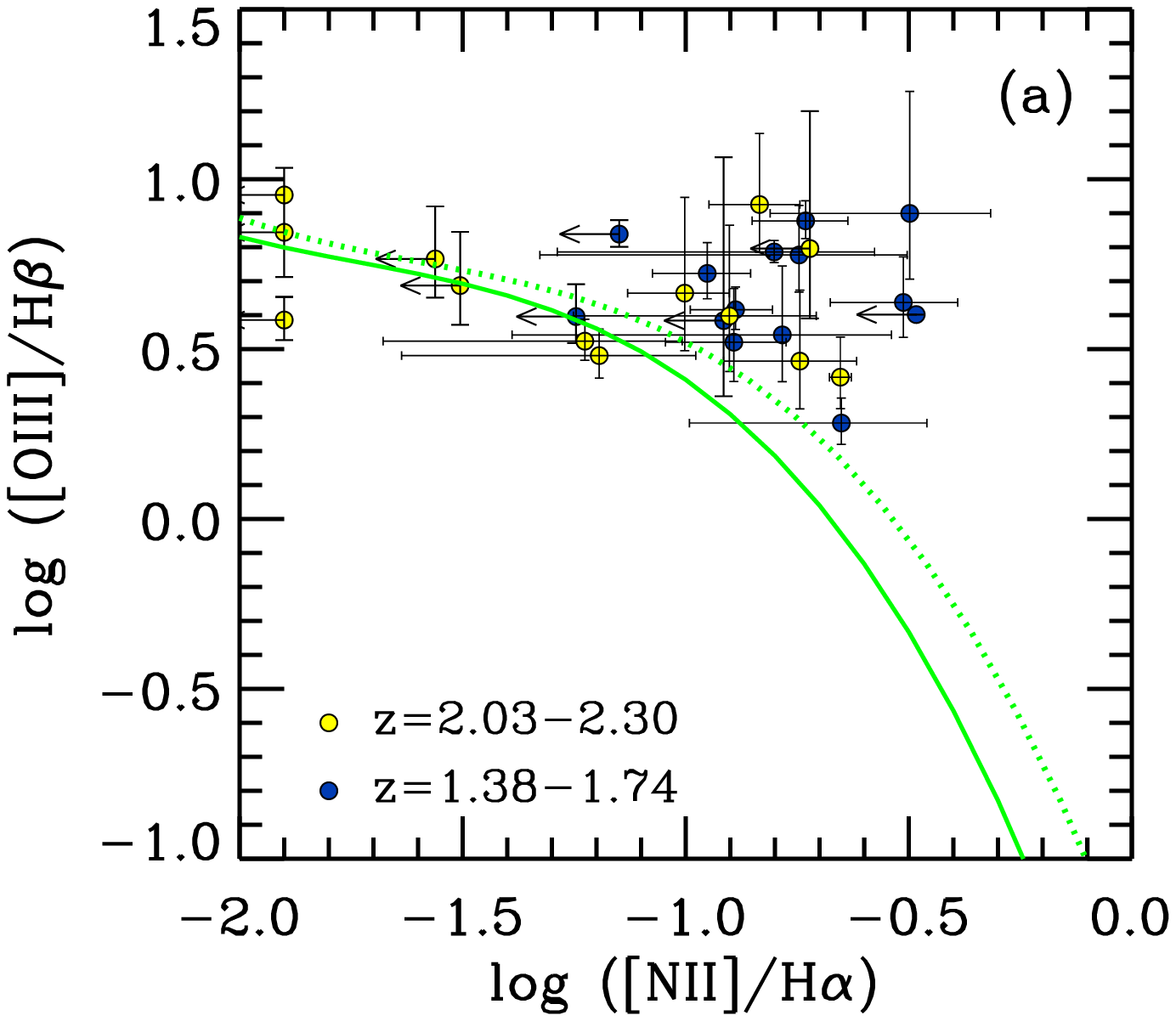}}
\centerline{\includegraphics[angle=0,width=6.0in]{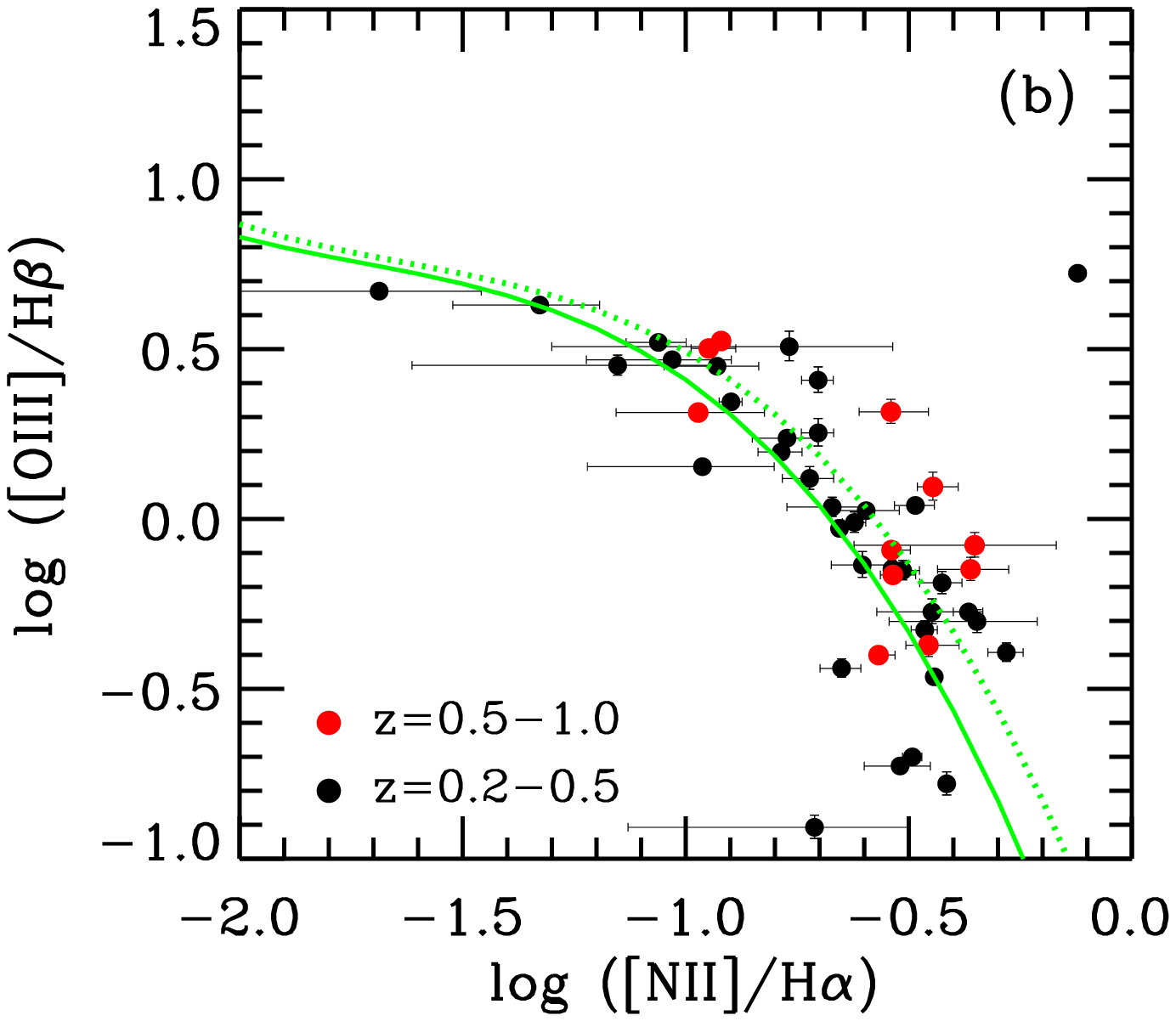}}
\caption{
Evolution of the BPT diagram to $z\sim2.3$ for galaxies in our sample with
(a) $\log L({\rm H}\beta)>41$~erg~s$^{-1}$ 
(blue circles---$z=1.38-1.74$) and $\log L({\rm H}\beta)>41.3$~erg~s$^{-1}$
(yellow circles---$z=2.03-2.30$) using the MOSFIRE data alone, and
(b) $\log L({\rm H}\beta)>40.5$~erg~s$^{-1}$ using either the DEIMOS data 
alone (black circles---$z = 0.2-0.5$) 
or the DEIMOS (O3Hb) plus MOSFIRE (N2Ha) data
(red circles---$z=0.5-1.0$). Source which are not detected
in N2Ha are shown as one sigma upper limits or at a nominal
value of N2Ha$=-1.9$.
The $1\sigma$ error bars on the ratios are larger for the MOSFIRE data 
than for the DEIMOS data. Note that some of the sources are likely AGNs,
given their locations on the diagram.
In both panels, the green solid curve shows the fit to the star-forming
galaxy locus of the full SDSS sample at the median 
$\log L({\rm H}\beta)=40.4$~erg~s$^{-1}$,
while the green dotted curves show the luminosity-offset loci obtained by plugging 
the median $\log L({\rm H}\beta)$ values of (a) 41.9~erg~s$^{-1}$ and 
(b) 41.5~erg~s$^{-1}$ 
into Equations~\ref{locus1} and \ref{locus2}, which vary N2Ha.
\label{bpt_goods_z}
}
\addtolength{\baselineskip}{10pt}
\end{figure*}

Having established the dependence of the BPT star-forming galaxy
locus on Balmer-line luminosity (but not redshift)  for the SDSS local sample, 
we turn our attention to the higher-redshift samples. 
As we did in Figure~\ref{sdss_bpt_highsfr_byz_points}, we start 
in Figure~\ref{all_bpt} by plotting
the SDSS data in two well-separated $\log L({\rm H}\beta)$
intervals, this time using
$39.0-40.0$~erg~s$^{-1}$ (black dots) and
$41.5-42.5$~erg~s$^{-1}$ (light green squares). The green curve is again 
the fit to the star-forming galaxy locus of the full SDSS sample
at the median $\log L({\rm H}\beta)=40.5$~erg~s$^{-1}$.
Compared to this curve, the lower-luminosity points lie
uniformly low, while the higher-luminosity points lie uniformly high.

Next, we plot our high-redshift galaxies, including every source with a 
well-measured ($>3\sigma$) O3Hb (purple circles). 
To increase the sample size, we also plot the high-redshift galaxies from
\citet[orange squares]{steidel14} and \citet[orange circles]{shapley15}.
We show galaxies with only upper limits on N2Ha at their 
$2\sigma$ limits (leftward pointing arrows). 

All three high-redshift galaxy samples overlap 
substantially with one another and are broadly consistent, though
our luminosity-selected sample extends to somewhat lower values of
N2Ha, while the mass-selected sample of 
\citet{shapley15} is slightly higher in N2Ha.
The \citet{steidel14} sample lies in between. 
Note that some of the sources are likely AGNs, given their locations 
on the diagram. 

The most striking observation from Figure~\ref{all_bpt} is how similar 
the high-redshift data are to the
higher-luminosity SDSS data for the 
star-forming galaxy locus. This emphasizes how critical it is to choose 
appropriate counterparts when comparing high-redshift samples with low-redshift ones.

The similarity of the high-redshift data to the higher-luminosity
SDSS data is not unique to the BPT diagram.
In Figure~\ref{alice_fig}, we show two
alternative diagnostic diagrams that are frequently used, 
(a) [NII]6584/[OII]3726,3729 (N2O2)
versus O3Hb and (b) [NII]6584/[SII]6717,6731 (N2S2) versus O3Hb. 
In both cases, the denominator is the sum of the fluxes from both
members of the doublet. For N2O2, we include
an extinction correction because of the wide wavelength separation of the lines. 
We plot the SDSS data in the same two well-separated
$\log L({\rm H}\beta)$ intervals as in Figure~\ref{all_bpt}.

We fitted the dependence of N2S2 on O3Hb and $L({\rm H}\beta)$ and
the dependence of N2O2 on O3Hb and $L({\rm H}\beta)$ for the 
SDSS star-forming galaxy locus (N2Ha~$<-0.5$) using the MPFIT2DFUN
two-dimensional fitting routine from \citet{markwardt09}. 
This time we adopted a
linear dependence for both O3Hb and $L({\rm H}\beta)$, obtaining
\begin{equation}
{\rm N2O2}=-0.71-0.75 ({\rm O3Hb}) + 0.16 (\log L({\rm H}\beta) - 40.5) 
\label{n2o2_o3hb}
\end{equation}
and
\begin{equation}
{\rm N2S2}=-0.20-0.40 ({\rm O3Hb}) + 0.17 (\log L({\rm H}\beta) - 40.5)\,.
\label{n2s2_o3hb}
\end{equation}
In Figure~\ref{alice_fig}, we show these linear fits at the median
$\log L({\rm H}\beta)=40.5$~erg~s$^{-1}$ (green line).
In both diagnostic diagrams, compared to this line,
the lower-luminosity points lie uniformly
low, while the higher-luminosity points lie uniformly high.
Because of the weakness of the [SII]6717,6731 lines and
the difficulty of both cross-calibrating the [OII]3726,3729
and [NII]6584 lines and applying appropriate extinction
corrections, we compare the SDSS data only to the deeper 
high-redshift galaxy sample of \citet[][orange circles]{shapley15},
who carefully treated the relative calibrations
of the [OII]3726,3729 and [NII]6584 lines. 
Again, the positions of the high-redshift galaxies in the diagrams
closely match the positions of the higher-luminosity SDSS galaxies.
The luminosities of the high-redshift galaxies in Figure~\ref{alice_fig}
lie between $\log L$(H$\beta$) = 41.1 and 42.6 with
85$\%$ lying in the 41.5--42.5 range \citep{reddy15} 
where we show the SDSS galaxies.

Finally, we consider the [OIII]5007,4861/[OII]3727 
versus ([OII]3727+[OIII]5007,4861)/H$\beta$ (O32 versus R23) diagram, which
is a well-known probe of both metallicity and ionization 
parameter \citep{kd02}. (Note that [OII]3727 is the
sum of the fluxes in the [OII] doublet.) 
In Figure~\ref{r23_o32}, we again show the higher-luminosity 
SDSS sample (light green squares) and the lower-luminosity SDSS 
sample (black points), together with the \citet{shapley15} high-redshift
data (orange circles). We include extinction corrections
in computing all of these ratios. 
The positions of the high-redshift galaxies in the diagram are fully consistent 
with the positions of the higher-luminosity SDSS 
galaxies, within the R23 errors (i.e., the high-redshift galaxies are 
more scattered due to the larger errors).

Now that we have established the similarity of the high-redshift data 
to the higher-luminosity SDSS data for all the diagnostic diagrams,
we return to the BPT diagram to examine in closer detail the
redshift evolution of our $L({\rm H}\beta)$-selected galaxy sample.
In Figure~\ref{bpt_goods_z}, we show 
the galaxies that satisfy the
luminosity limits at which our samples are complete 
(see Section~\ref{grisdata}) for
(a) two high-redshift and (b) two low-redshift intervals.
In (a), we show the galaxies 
with $\log L({\rm H}\beta)>41.0$~erg~s$^{-1}$
and $z = 1.38-1.74$ as blue circles
and the galaxies with $\log L({\rm H}\beta)>41.3$~erg~s$^{-1}$ 
and $z = 2.03-2.30$ as purple circles. The median luminosities are 
$\log L({\rm H}\beta)=41.9$~erg~s$^{-1}$ for both samples.
We made these measurements solely from the MOSFIRE data.
In (b), we show galaxies with 
$\log L({\rm H}\beta)>40.5$~erg~s$^{-1}$ and either $z = 0.2-0.5$ 
(black circles) or $z=0.5-1$ (red circles). 
Here the median luminosities are $\log L({\rm H}\beta)=40.9$~erg~s$^{-1}$
and 41.5~erg~s$^{-1}$, respectively. We made the $z=0.2-0.5$ measurements 
solely from the DEIMOS data, while we made the $z=0.5-1.0$
measurements from both the DEIMOS (O3Hb) and MOSFIRE (N2Ha) data. 
We have included $1\sigma$ error bars on all the data points based on randomized
measurements in the spectra. The error bars are generally very small for O3Hb
in the DEIMOS data but much larger in the MOSFIRE data. 
Note that some of the data points lie off the locus of star-forming
galaxies, in the regime consistent with being powered by AGNs.

In both panels, we show the fit to the star-forming galaxy locus 
of the full SDSS sample (N2Ha~$<-0.5$) at the median 
$\log L({\rm H}\beta)=40.0$ (green solid curve), along with the
luminosity-offset locus determined by evaluating 
Equations~\ref{locus1} and \ref{locus2}, which 
vary N2Ha (green dotted), at median $\log L({\rm H}\beta)$ values of
(a) $41.9$~erg~s$^{-1}$ and 
(b) $41.5$~erg~s$^{-1}$ 
(using the median for the $z=0.5-1$ interval for (b)).
As already expected from Figure~\ref{all_bpt}, we see that we can
smoothly represent our high-redshift data with increasing luminosity 
based on the luminosity offset calculated from the SDSS data.    
Put simply, galaxies of a given $L({\rm H}\beta)$ have the same strong
emission-line properties, regardless of redshift.


\section{Luminosity-Adjusted Metallicity Relations for the SDSS Galaxies}
\label{secdisc}

A major goal of this paper is to explore what underlying properties in the galaxies 
might be causing the observed offsets with $\log L({\rm H}\beta)$
in the diagnostic diagrams. 
We do this by examining the various line ratios individually.
For reference, the recent compilation of \cite{asplund09}
gives  solar abundances of
$12+\log$(O/H)$=8.69$, 
$12+\log$(N/H)$=7.83$ and $12+\log$(S/H)$=7.12$.
The  models of \citet{kd02} discussed below use the
values from \citet{anders89}: 
$12+\log$(O/H)$=8.93$,
$12+\log$(N/H)$=8.05$ and $12+\log$(S/H)$=7.21$, together
with dust depletions of O and N of $-$0.22 dex.

There are several well-known mechanisms that can produce offsets
in the diagnostic diagrams, including changes in the ionization
parameter (i.e., $q$, the ratio of the ionizing
photon flux to the hydrogen number density), 
relative abundance variations (including the effects
of dust depletion), particularly in the
N to O ratio, and changes in the hardness of the ionizing spectrum.
If the underlying mechanism(s) are similar in the SDSS and high-redshift
samples, then we can parameterize
the dependence of the metallicity relations on
$\log L({\rm H}\beta)$ using the SDSS sample and generate 
luminosity-adjusted metallicity relations that may be applicable 
to the high-redshift samples. However, regardless of whether
these relations really apply at high redshift, the addition of the second parameter (i.e., line luminosity) tightens
the relations and improves the low-redshift metallicity estimates.

As we shall discuss further below, the N2O2
parameter is insensitive to the ionization and therefore is primarily
a measure of the relative N/O abundance \citep[e.g.][]{perez09}.
The high N2O2 ratio in the high-redshift galaxies has been interpreted
as showing that N/O is overabundant in these objects
\citep{masters14,steidel14,shapley15}. 
\citet{perez09} give an empirical calibration based on local galaxies of

\begin{equation}
{\rm log (N/O)}= 0.93 ({\rm N2O2}) - 0.20 
\label{n2o2cal}
\end{equation}
 
\noindent which, when combined with Equation~\ref{n2o2_o3hb}, gives a
relationship for the median track at a given $\log$ L(H$\beta$) of

\begin{equation}
{\rm log(N/O)}= -0.86 -0.70 ({\rm O3Hb}) + 0.15 (\log L({\rm H}\beta) - 40.5) 
\label{n2o2_no}
\end{equation}

For O3Hb=0.00, $\log$(N/O) would range from $-$1.08 at $\log L$(H$\beta$) = 39 to
$-$0.64 at $\log L$(H$\beta$) = 42, compared to the adopted solar value of $-$0.86,
while at O3Hb = 0.50 the range would be $-$1.43 to $-$0.98. However, the
uncertainties in the absolute values should be borne in mind. More
robustly, we can see from Equation~\ref{n2o2_no} that at a fixed
value of O3Hb, high-luminosity objects ($\log L$(H$\beta$) = 42)
would have a N/O abundance
which is a factor of 1.7 times higher than the average local
galaxy ($\log L$(H$\beta$) = 40.5); this is the same effect 
that is seen in the high-redshift samples.

Absent any other effects,
the increase in N/O with increasing $\log$ L(H$\beta$) will
produce an increase of N2Ha at a fixed O3Hb, resulting
in a shift in the BPT diagram in the sense that is seen
in Figure~\ref{sdss_bpt_by_sfr}. However the dependence
on $\log$ L(H$\beta$) is too strong, with a slope of 0.15
in Equation~\ref{n2o2_no} compared with 0.09 in the
evolution of the BPT diagram (Equations~\ref{locus1}-\ref{locus3}).
This suggests that there must be a countervailing
trend that reduces N2Ha for a given O3Hb and reduces
the evolution in the BPT diagram.

\begin{figure*}
{\includegraphics[angle=0,width=3.2in]{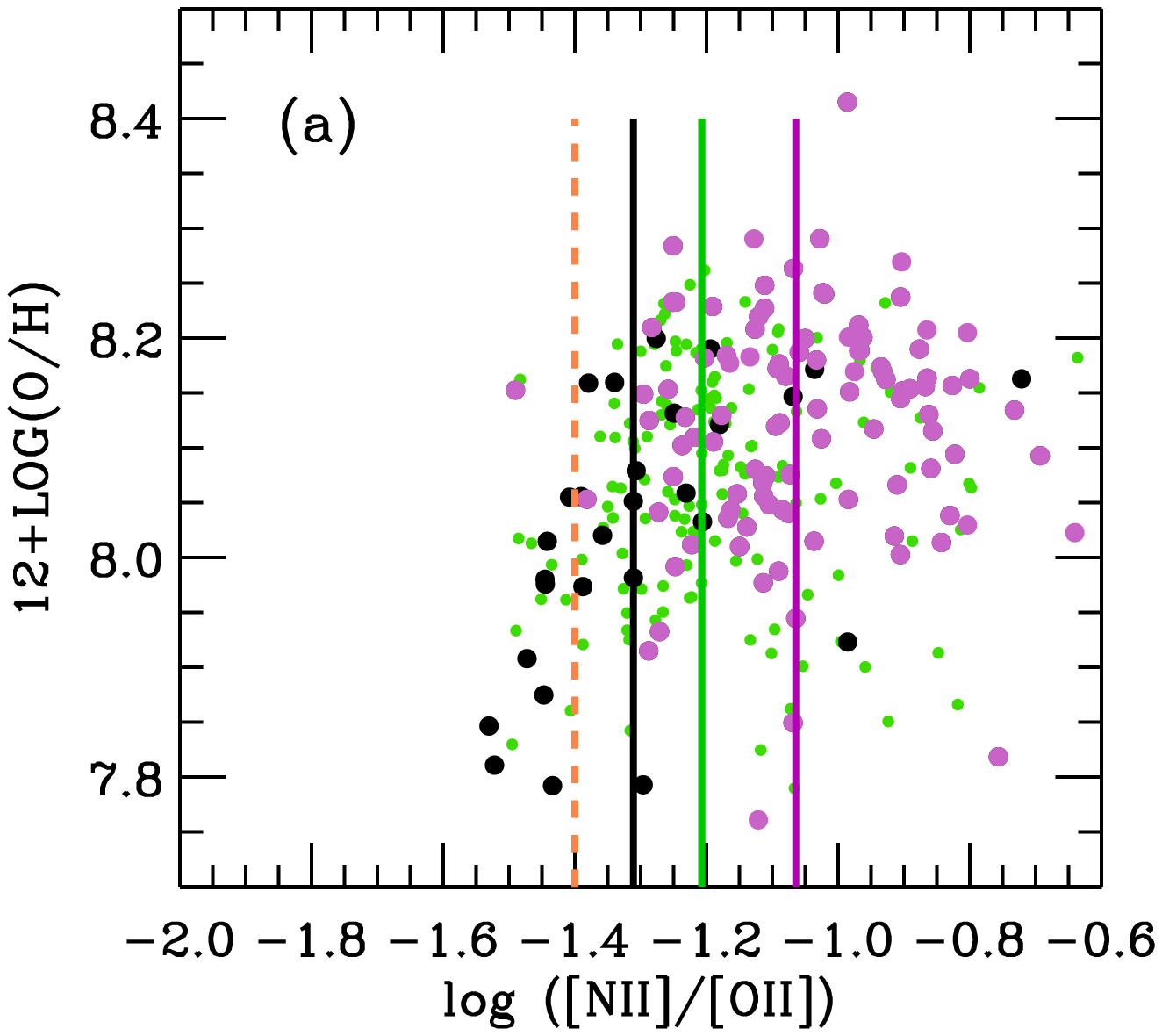}}
{\includegraphics[angle=0,width=3.2in]{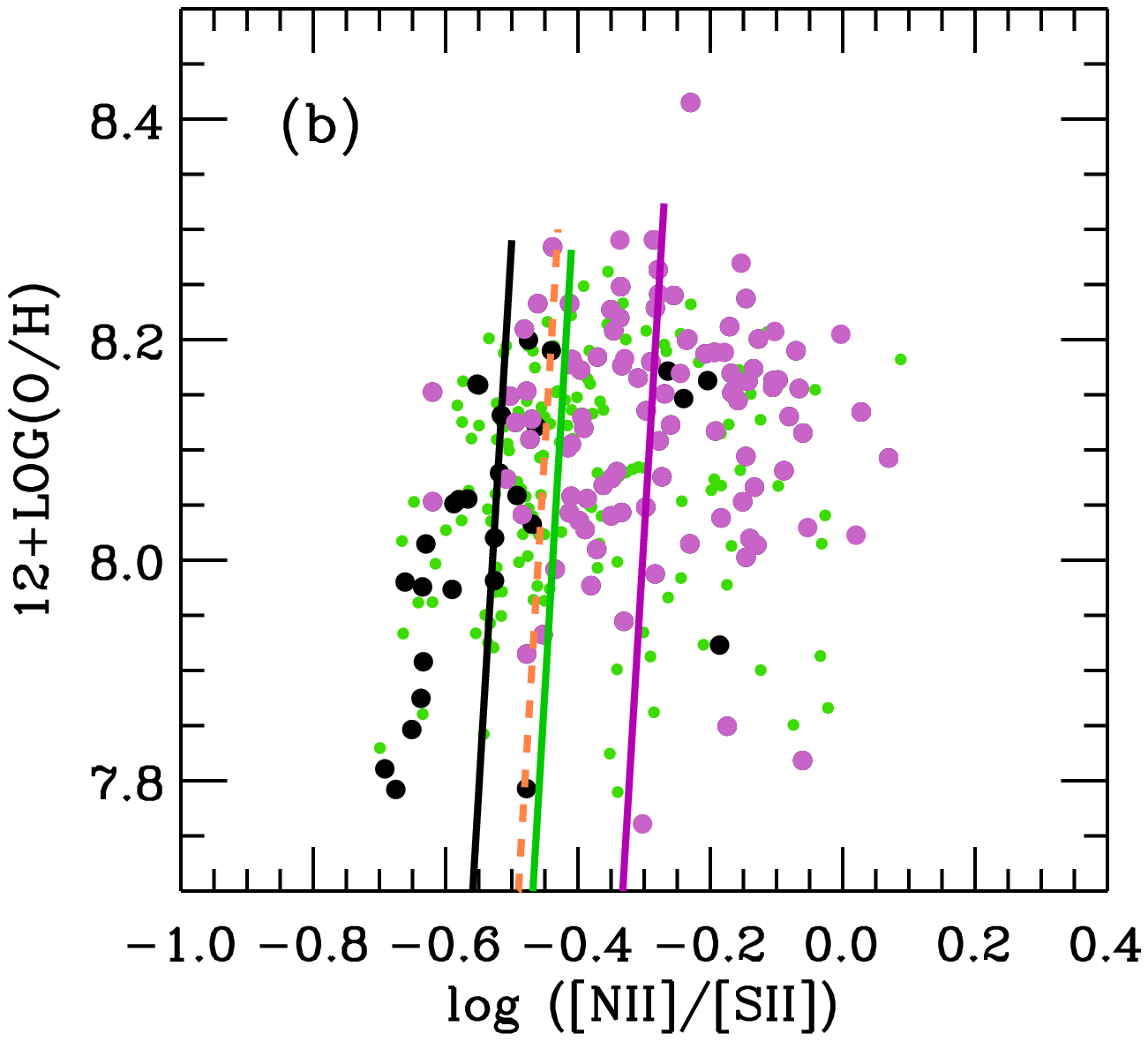}}
{\includegraphics[angle=0,width=3.2in]{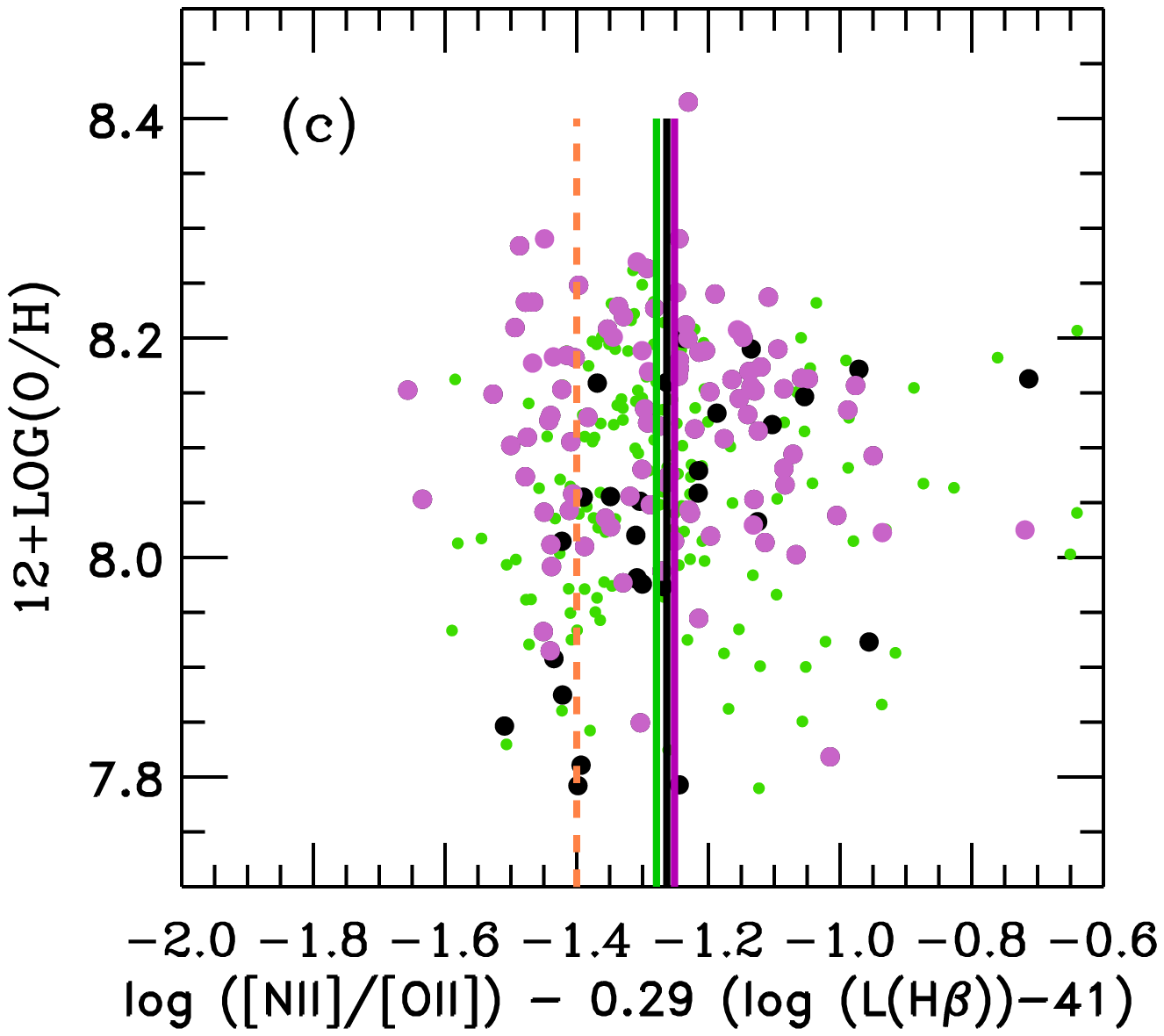}}
\hspace*{1.75cm}{\includegraphics[angle=0,width=3.2in]{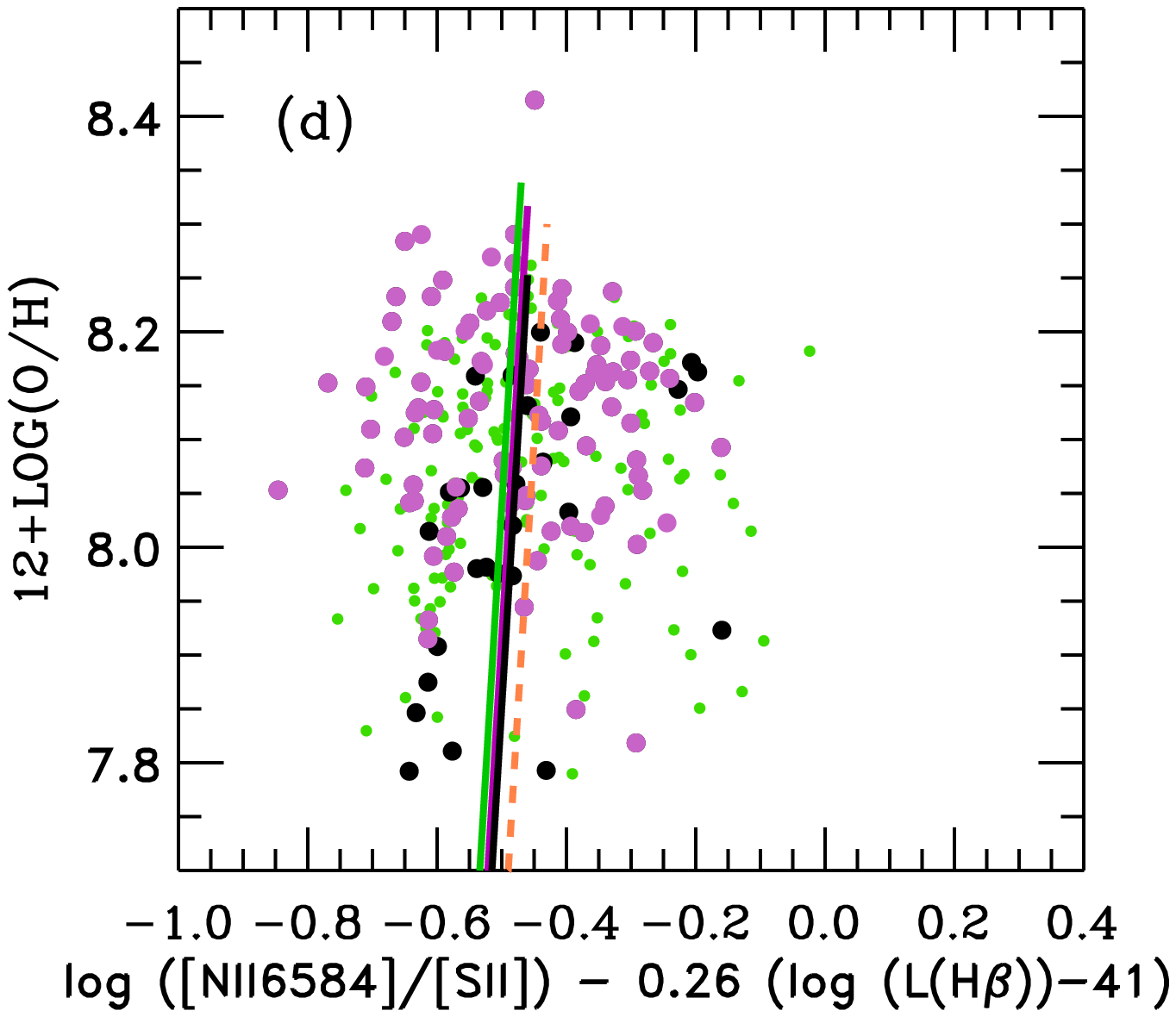}}
\caption{Direct O abundance measurements  vs.\ (a) N2O2 and (b) N2S2 for galaxies 
in the full SDSS sample that lie on the star-forming galaxy locus of the BPT diagram 
(N2Ha~$<-0.5$) and have strong ($>10\sigma$) [OIII]4363 detections,
plotted in three intervals of $\log L({\rm H}\beta)$
(black circles---$40.5-41.0$~erg~s$^{-1}$; green circles---$41.0-41.5$~erg~s$^{-1}$;
purple circles---$41.5-42.0$~erg~s$^{-1}$).
Our fits to the theoretical values calculated in \cite{kd02} are
shown as the orange dashed lines. Assuming these shapes, we computed
offsets to the median values of the colored symbols
(corresponding colored lines).
Also shown are directly measured O abundances vs.
(c) N2O2$-0.29(\log L({\rm H}\beta) - 41)$ and
(d) N2S2$-0.26(\log L({\rm H}\beta) - 41)$,
which are the luminosity-adjusted values required to bring the samples 
into alignment.
\label{lhb_others}
}
\addtolength{\baselineskip}{10pt}
\end{figure*}

\begin{figure*}
\hspace*{-0.6cm}{\includegraphics[angle=0,width=3.0in]{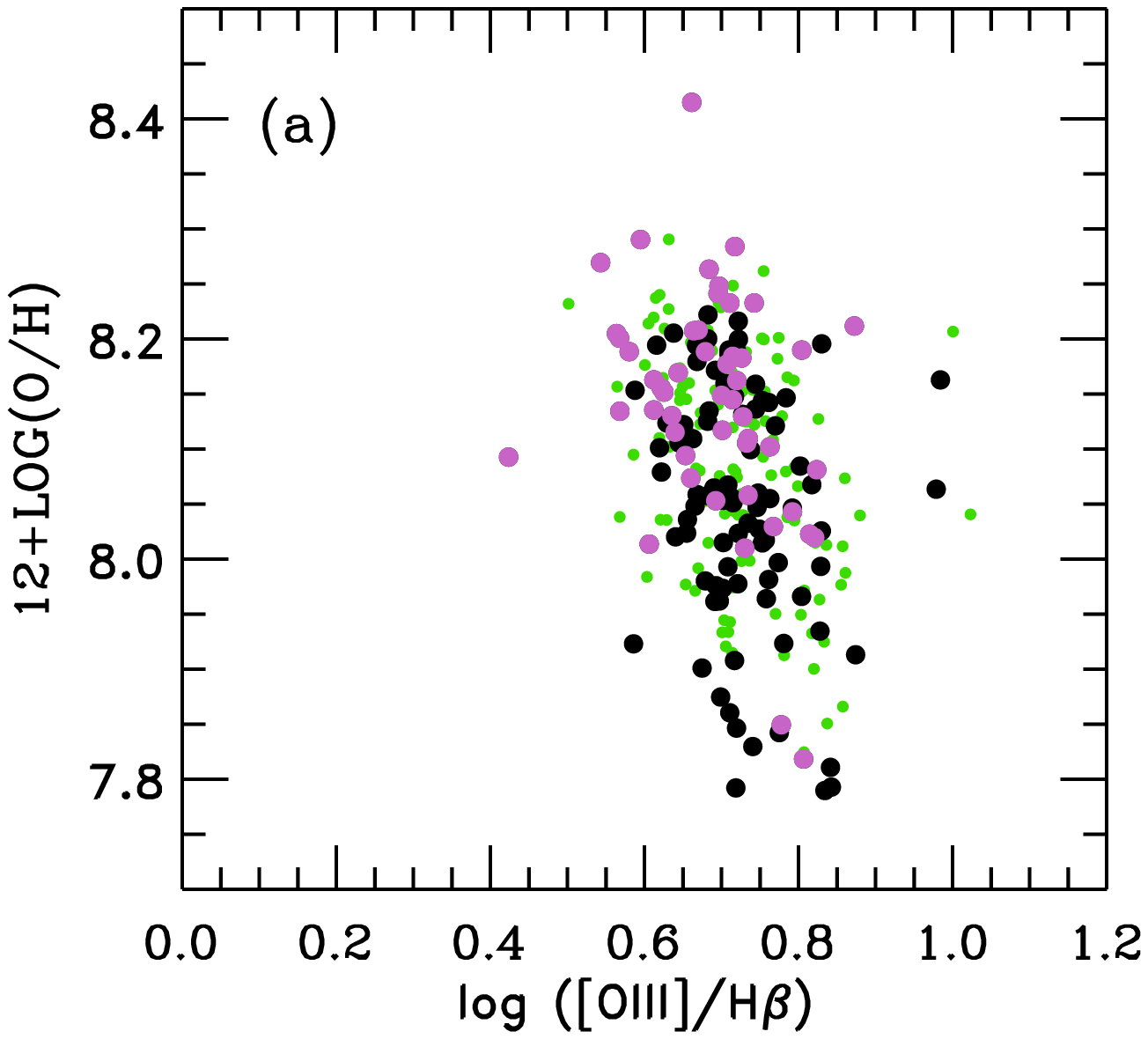}}
\hspace*{-1.85cm}{\includegraphics[angle=0,width=3.0in]{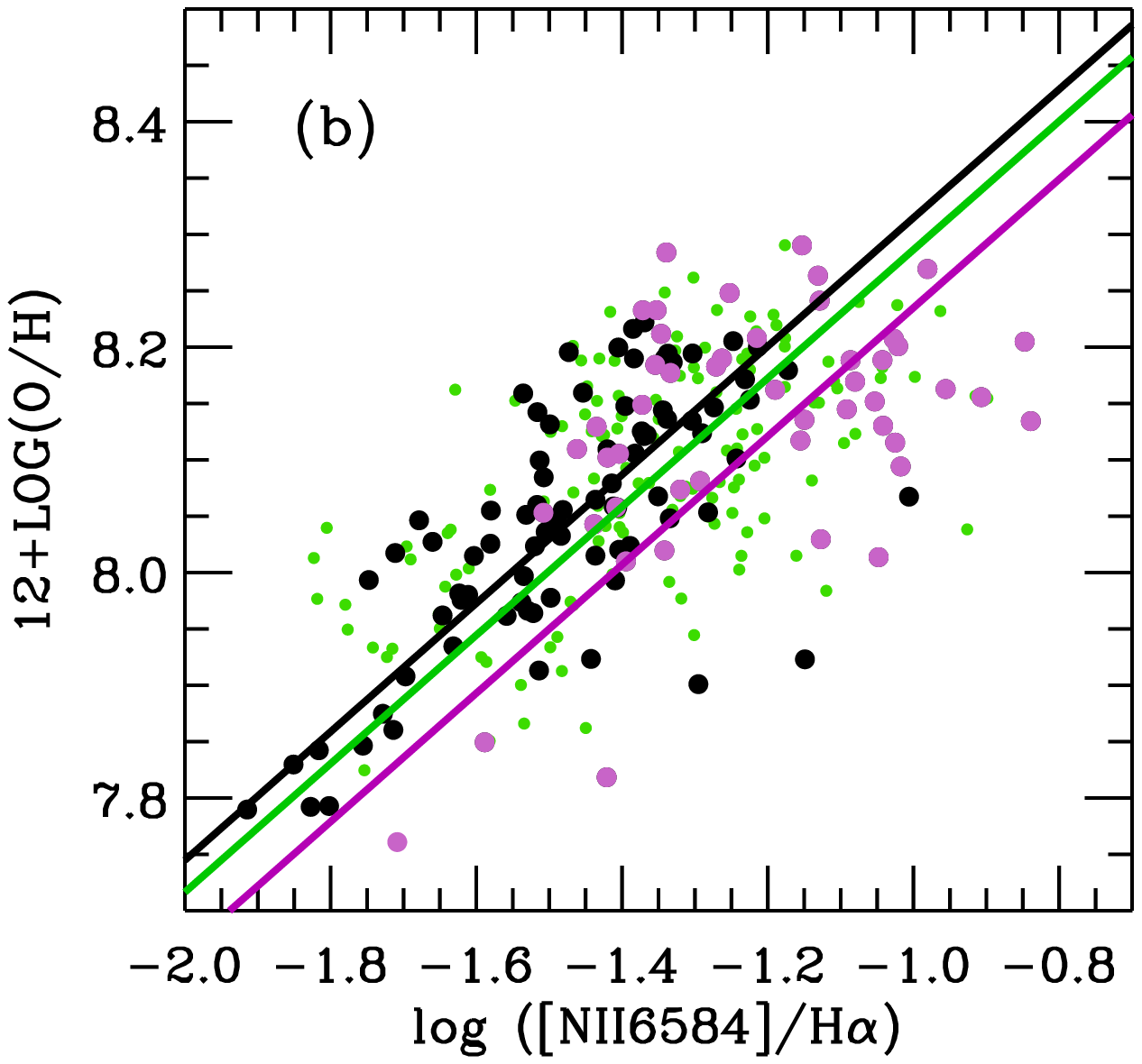}}
\hspace{-1.65cm}{\includegraphics[angle=0,width=3.0in]{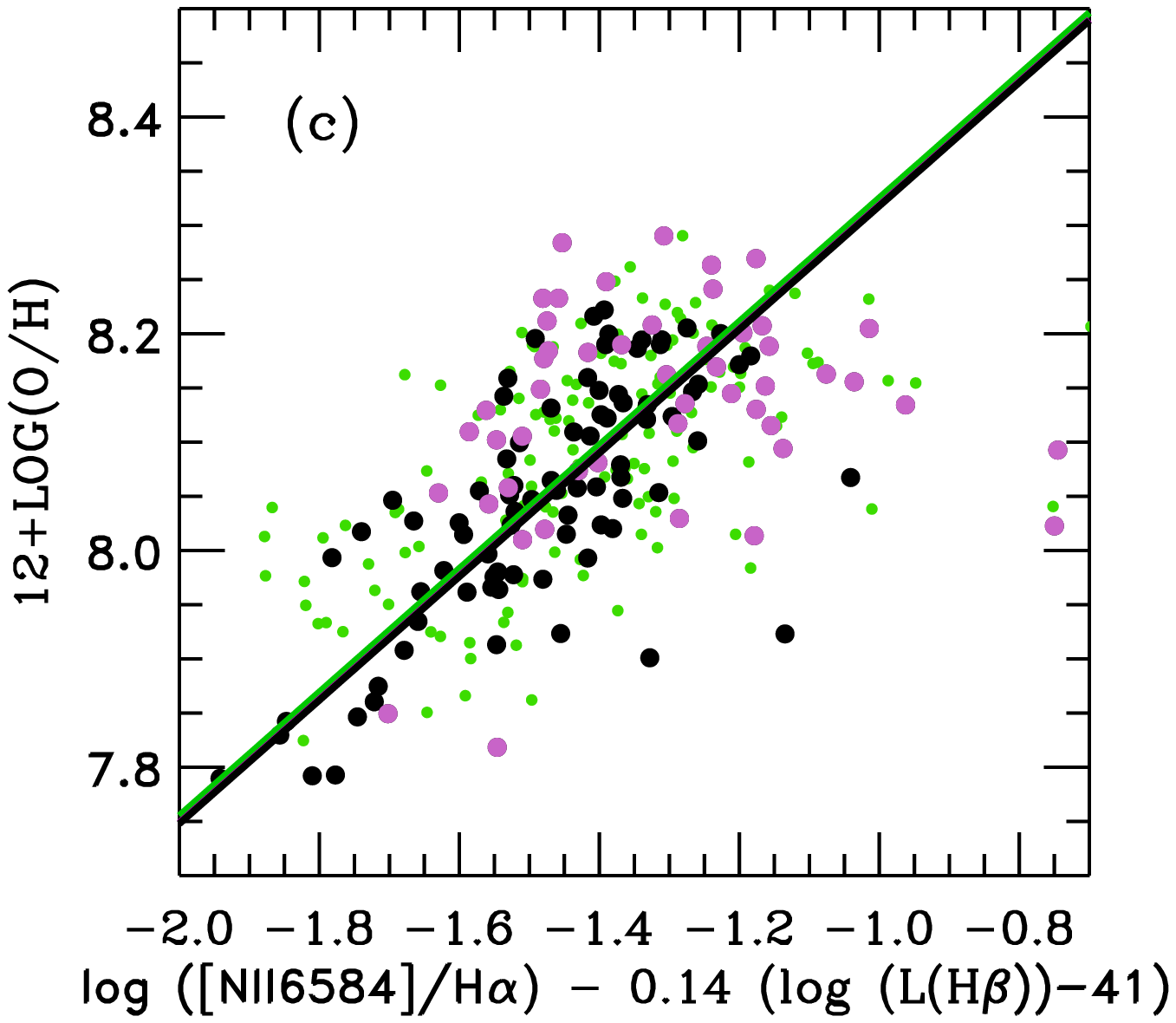}}
\caption{(a) Direct O abundance measurements  vs.\ O3Hb for galaxies in
the full SDSS sample that lie on the star-forming galaxy locus of the BPT
diagram (N2Ha~$<-0.5$) and have strong ($>10\sigma$) [OIII]4363 detections,
plotted in three intervals of $\log L({\rm H}\beta)$
(black circles---$40.5-41.25$~erg~s$^{-1}$; green 
circles---$41.25-41.75$~erg~s$^{-1}$;
purple circles---$41.75-42.5$~erg~s$^{-1}$).
For the same galaxies, direct O abundance measurements  vs.\ (b) N2Ha and
(c) N2Ha$-0.14\log L({\rm H}\beta)$ 
(``adjusted'' N2Ha). 
The colored lines have the \cite{pp04} slope and
match the median values of the corresponding colored symbols.
\label{lhb_n2ha}
}
\addtolength{\baselineskip}{10pt}
\end{figure*}

We investigate this issue  by comparing the 
strong emission-line ratios with direct O abundances,
which we calculate using
the relations defined in \cite{izotov06}. We use only SDSS galaxies
on the star-forming galaxy locus of the BPT diagram (N2Ha~$<-0.5$)
where the [OIII]4363 line is strongly detected ($>10\sigma$). Unfortunately,
this strong detection requirement will result in a
 bias toward lower metallicity  
(where the [OIII]4363 line is stronger) and higher Balmer-line luminosity, 
which means we will have a relatively limited dynamic range for determining 
dependences on $\log L({\rm H}\beta$).

\subsection{N2O2 and N2S2}
\label{secn2o2}

We first consider the changes in N2O2 and N2S2. Both these ratios 
are extremely weakly dependent on $q$.  
At low O abundances relative to solar 
these line ratios are also almost independent of metallicity
\citep[see, e.g. Figures~3 and 4 of][]{kd02}.

In Figure~\ref{lhb_others}, we show $12+\log$(O/H) 
vs.\  (a) N2O2 and (b) N2S2. 
We use colored circles to denote three 
intervals of $\log L({\rm H}\beta)$ (black---$40.5-41.0$~erg~s$^{-1}$;
green---$41.0-41.5$~erg~s$^{-1}$, and purple---$41.5-42.0$~erg~s$^{-1}$).
Over the range $12+\log$(O/H)~$<8.4$, we assume
that N2O2 is independent of $12+\log$(O/H), and
over the range $12+\log$(O/H)~$<8.3$, we use a linear 
fit to the extremely weak dependence of N2S2 on $12+\log$(O/H)
from \cite{kd02} (orange dashed lines).
We use these slopes to compute the offsets to the medians of the 
colored symbols (colored lines), but we do not find the offsets 
between the luminosity samples to be sensitive to the adopted slopes.

We find that the different luminosity samples can 
be brought into consistency using offsets that are linear functions of 
$\log L({\rm H}\beta)$. In Figure~\ref{lhb_others}(c), we show
$12+\log$(O/H)  vs.\ N2O2$-0.29(\log L({\rm H}\beta)-41)$,
and in Figure~\ref{lhb_others}(d), we show
$12+\log$(O/H)  vs.\ N2S2$-0.26(\log L({\rm H}\beta)-41)$.
We have checked that the offsets are not sensitive to the choice of
the N2Ha limit used to define the star-forming galaxy locus in the BPT 
diagram, nor to the use of a luminosity-adjusted N2Ha limit
(Equations~\ref{locus1} and \ref{locus2}). We also checked that
the offsets are not sensitive to the choice of S/N in the [OIII]4363 line 
detection; we obtained a similar result using a $5\sigma$ 
threshold instead of a $10\sigma$ threshold.

As we have discussed above the offsets in the N2O2 and N2S2 ratios
must be understood as arising from changes in the gas-phase abundance 
ratios of N relative to O and N relative to S, with the 
higher-luminosity sources 
being progressively more abundant in N relative to O, and
the O and S abundances not changing relative to each other.
Since changes in the dust depletion would result in 
S (which does not deplete onto dust) changing relative to O,
it is likely that the changes are in the overall
abundances (i.e., gas plus dust) rather than only in the gas-phase abundances.

More specifically, we can interpret 
Figures~\ref{lhb_others}(c) and \ref{lhb_others}(d) 
as implying that the average relations beteween
log(N/O) and log(N/S) and log L(H$\beta$) for these
luminous galaxies are 

\begin{equation}
{\rm log(N/O)} = -0.74+0.29(\log L({\rm H}\beta)-41) \,,
\label{abund1}
\end{equation}
and 
\begin{equation}
{\rm log(N/S)} =  0.64+0.26(\log L({\rm H}\beta)-41) \,.
\label{abund2}
\end{equation}
The normalizations in these equations are chosen to match
the observed means shown in Figure~\ref{lhb_others} to
the \citet{kd02} 
models shown by the orange dashed
lines.

Relative to the local solar value of $\log({\rm N/O})=-0.86$, 
Equation~\ref{abund1} would imply that the high-$L({\rm H}\beta)$ galaxies 
(and hence the
high-redshift galaxies) are, on average,
over-abundant in N relative to O by a factor of 2.2
at $\log L({\rm H}\beta)=41.6$~erg~s$^{-1}$.

The above $L({\rm H}\beta)$ dependences derived from the direct O abundance 
measurements are slightly steeper than the $L({\rm H}\beta)$ dependences of
Equations~\ref{n2o2_o3hb}, \ref{n2s2_o3hb}  and \ref{n2o2_no}
for N2O2 and N2S2
versus O3Hb. This does not appear to come from variations of
O3Hb with luminosity, since, for the observed range of O3Hb,
 we see no dependence of O3Hb
on $L({\rm H}\beta)$ for the galaxies with direct [OIII]4363 measurements.
We illustrate this in Figure~\ref{lhb_n2ha}(a), where we show
$12+\log$(O/H) versus O3Hb for the same
three intervals of $\log L({\rm H}\beta)$ used in Figure~\ref{lhb_others}.
The lack of dependence on $\log L({\rm H}\beta)$
is a consequence of the low metallicities and high ionization parameters in
the sample with direct O measurements, which
result in low values of [OII]/[OIII], and values of [OIII]5007/H$\beta$ that
are insensitive to the ionization parameter and metallicity.

Instead, the $L({\rm H}\beta)$ dependences derived from the direct O abundance  
measurements reflect the bias of galaxies with such measurements 
toward a more limited range in $L({\rm H}\beta)$. When we limit
the N2O2 and N2S2 vs.\ O3Hb fits to the same $L({\rm H}\beta)$
range, we find similarly steep relations.

To summarize, our results show that it is the increases in the
overall abundance ratios of N/O 
and N/S that are producing the offsets seen in the N2O2 and N2S2 
diagnostic diagrams of 
Figure~\ref{alice_fig} for the higher-luminosity SDSS
sample. The high-redshift samples are similar in over-abundance
to the low-redshift samples with the same high $L$(H$\beta$).

\subsection{N2Ha}
\label{secn2ha}

We now examine the change in N2Ha and its role
in the observed evolution  of the BPT diagram.
To do so, we start by considering the \cite{pp04} N2Ha calibrator
(hereafter, Pettini-Pagel relation), which has
been widely used to measure metallicities
in high-redshift galaxies, though with some concerns 
\citep[e.g.,][]{erb06,liu08,newman14,sanders15}.  
While empirically the relation between direct O abundance measurements 
and N2Ha is good at low metallicities \citep{pp04,maiolino08},
the relation has always been treated with some suspicion because
it is strongly dependent on both the ionization parameter and the 
abundance ratio of N/O.

Not surprisingly, there is also a dependence of the Pettini-Pagel relation 
on $L({\rm H}\beta)$. In Figure~\ref{lhb_n2ha}(b), we show 
$12+\log$(O/H) versus N2Ha for the same three intervals of 
$\log L({\rm H}\beta)$ used in Figure~\ref{lhb_others}.
Adopting the slope of 0.57 used by \cite{pp04} for their calibration, 
we match the medians of the colored symbols between $-2$ and
$-1.3$ (colored lines). We see that the higher-$L({\rm H}\beta)$ 
galaxies have a higher N2Ha for a given direct O abundance.
We remove this luminosity dependence in Figure~\ref{lhb_n2ha}(c)
by plotting $\log L({\rm H}\beta)$ vs.
N2Ha$-0.14(\log L({\rm H}\beta) -41)$.
This gives a luminosity-adjusted relation of
\begin{eqnarray}
{\rm 12+\log(O/H)} = 8.90 &+& 0.57 ({\rm N2Ha}) \nonumber \\ &-& 0.14{(\log L({\rm H}\beta)-41)} \,.
\label{ppeqn}
\end{eqnarray}
The luminosity correction is large enough to have a significant
effect on metal measurements in high-redshift galaxies, which generally
will be over-estimated by about 0.1~dex using the uncorrected relation.

The N2Ha dependence on $L({\rm H}\beta)$ is a combination of 
two effects: the N/O ratio increasing with
luminosity, and the ionization parameter increasing with
luminosity. These drive the relation between $12+\log$(O/H)
and N2Ha in opposite directions, with increasing N/O
raising N2Ha for a given O abundance, while increasing
$q$ reduces it.

To disentangle the two effects,
we first consider how $q$ depends on $L({\rm H}\beta)$.
We calculate $q$ using the \citet{kd02}  [OIII]/[OII]$-q$ relation,
the parameterization for which is given in Equation~13 of \citet{kk04}. 
In cases where the [OIII]4363 line is detected above the $10\sigma$
level, we again make direct O abundance measurements using
the relations defined in \cite{izotov06}. Otherwise, we use the 
strong-line-based O abundances 
given in the value-added SDSS catalog described
in Section~\ref{sdssdata}, which are calculated
using the methods of \cite{tremonti04}.

\begin{inlinefigure}
\hspace*{-0.4cm}\includegraphics[angle=0,width=4.0in]{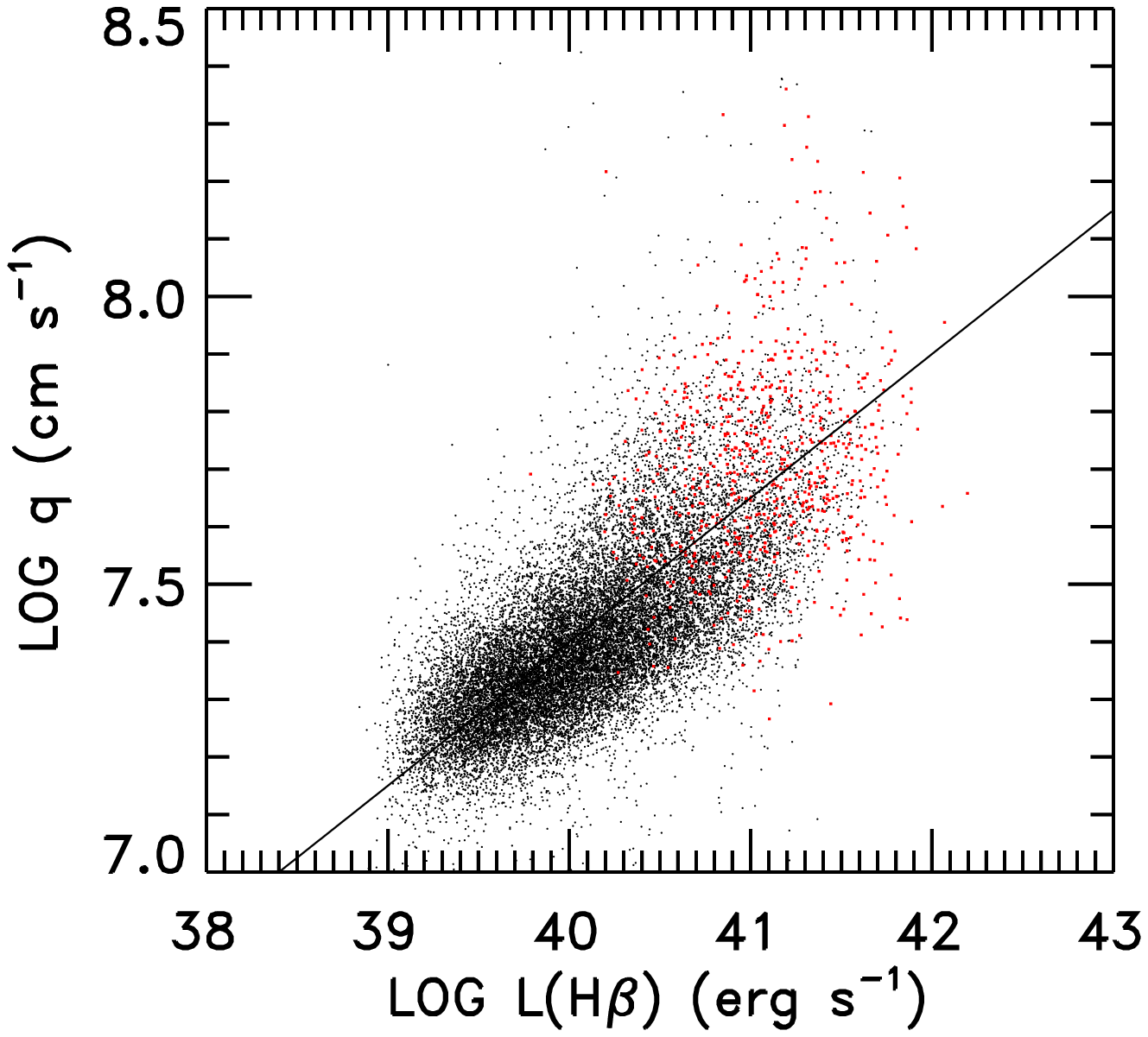}
\caption{Logarithmic ionization parameter vs.
$\log L({\rm H}\beta)$. Red dots show where the O abundances---on
which the measurement of $q$ is based---were determined using
the direct method, while the black dots show where the O abundances
were determined using the \cite{tremonti04} strong-line methodology. 
The solid line is a linear fit to the data.
\label{ionization}
}
\addtolength{\baselineskip}{10pt}
\end{inlinefigure}

As we show in Figure~\ref{ionization}, there is quite a tight relation between
$q$ and $\log L({\rm H}\beta)$. We approximate this with
\begin{eqnarray}
\log q = 7.51 + 0.21{(\log L({\rm H}\beta)-41)} \,,
\label{ioneqn}
\end{eqnarray}
which we show as the solid line.
The galaxies with direct O abundance measurements (red dots) lie primarily
in the high-$L({\rm H}\beta)$ regime. This is partly a consequence
of higher-luminosity galaxies having higher ionization parameters, and
partly a consequence of the selection bias introduced
by the higher line S/N in these spectra, since the brighter
the galaxy, the more likely it will satisfy the S/N criterion
for measuring [OIII]4363. It is reasonable to ask whether using the 
strong-line-based
O abundances introduces any biases in Equation~\ref{ioneqn}.
We have therefore repeated the fit using only the 
strong-line-based O abundances, but we find
essentially the same result.

The range in $\log q$ is relatively modest, from
7.2 at low $L({\rm H}\beta)$ to 7.9 at high $L({\rm H}\beta)$, 
or roughly a multiplicative factor of 4 in $q$. As we 
noted in \citet{cb08}, it is for this reason that even
ionization-sensitive diagnostics, such as N2Ha, can
be reasonably good metallicity estimators.
However, the close dependence of $q$ on $\log L({\rm H}\beta)$ 
indicates that the metallicity calibration relations
can be improved by including $L({\rm H}\beta)$ as a 
second parameter.

To determine the effects of the dependence of
N2Ha on $q$, we make use of the
\citet[their Figure~7]{kd02} model calculations, which we show
in Figure~\ref{n2ha_q}(a) for the range of $\log q$ of
interest ($\sim7.3-8.2$).
For this limited range, we can bring the N2Ha values
into consistency by applying a simple offset of
$0.68(\log q-7.5)$. We illustrate this in Figure~\ref{n2ha_q}(b),
where we apply this offset to the models. 
Below $12+\log$(O/H)=9, the results are well fit by a linear relation (black line),
\begin{eqnarray}
12+\log {\rm (O/H)} = 9.29 &+& 0.82\log ({\rm N2Ha}) \nonumber \\ &+& 0.68(\log q - 7.5) \,.
\label{abundance}
\end{eqnarray}
Combining Equations~\ref{ioneqn} and  \ref{abundance} gives
\begin{eqnarray}
12+\log {\rm (O/H)} &=& 9.35 + 0.82\log ({\rm N2Ha}) \nonumber \\ &+& 0.14(\log L({\rm H}\beta) - 41) \,.
\label{lhb_abundance}
\end{eqnarray}
This has the opposite sign in the dependence on $L({\rm H}\beta)$
from the relation derived from the observational data (Equation~\ref{ppeqn}).
However, when we allow for the variation in N/O abundance with
$L({\rm H}\beta)$ described by Equation~\ref{abund1}, the dependence
reverses sign, and Equation ~\ref{lhb_abundance} becomes
\begin{eqnarray}
12+\log {\rm (O/H)} &=& 9.20 + 0.82\log ({\rm N2Ha}) \nonumber \\ &-& 0.10(\log L(H\beta) - 41) \,,
\label{lhb_abundance_final}
\end{eqnarray}
which is fully consistent with the observational dependence on $L({\rm H}\beta)$.
Equation~\ref{lhb_abundance_final} has a slightly
steeper dependence on N2Ha than the Pettini-Pagel
relation of Equation~\ref{ppeqn}, but it is in close numerical agreement
over most of the fitted range.

The ionization dependence also explains why the evolution in the
BPT diagram is weaker than that which would be produced by the
evolution in the N/O ratio with $\log L$(H$\beta$)
(Equation~\ref{abund1}). The increase
in $q$ reduces the [NII]6584/[OIII]5007 ratio and offsets the
rise in N2Ha for a given O3Hb produced by the increase in N/O.
For $12+\log$(O/H) in the range 8.5 to 9 and $\log q$ in the range
7.2 to 7.8, Figure~8 of \cite{kd02} shows [NII]6584/[OIII]5007
changing as $-1.2 \log q$. Combining with Equation~\ref{ioneqn}
translates this to [NII]6584/[OIII]5007 changing as
$-0.25 \log L$(H$\beta$). Combining the ionization reduction with
the N/O increase gives N2Ha increasing as $0.04 \log L$(H$\beta$)
for a given O3Hb. While this is weaker than the measured value of
$0.09 \log L$(H$\beta$) in Equation~\ref{locus1}, it is well within
the uncertainties in the calculations. It therefore appears that
the offsets in the BPT diagram can be understood
as being caused by changes in the N/O abundances, which are partially
offset by the changes in the ionization parameter.


\begin{inlinefigure}
\centerline{\includegraphics[angle=0,width=4.0in]{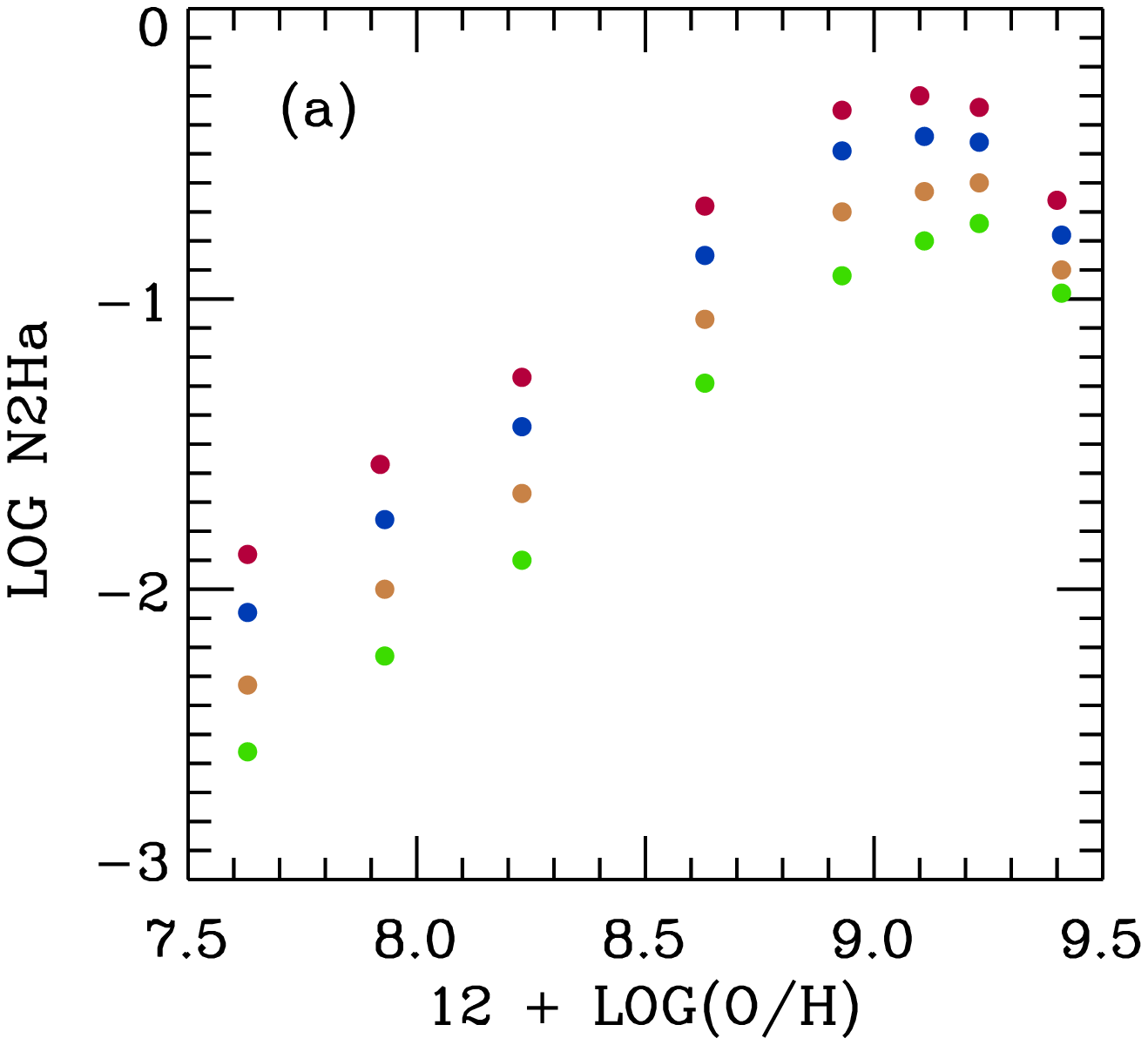}}
\centerline{\includegraphics[angle=0,width=4.0in]{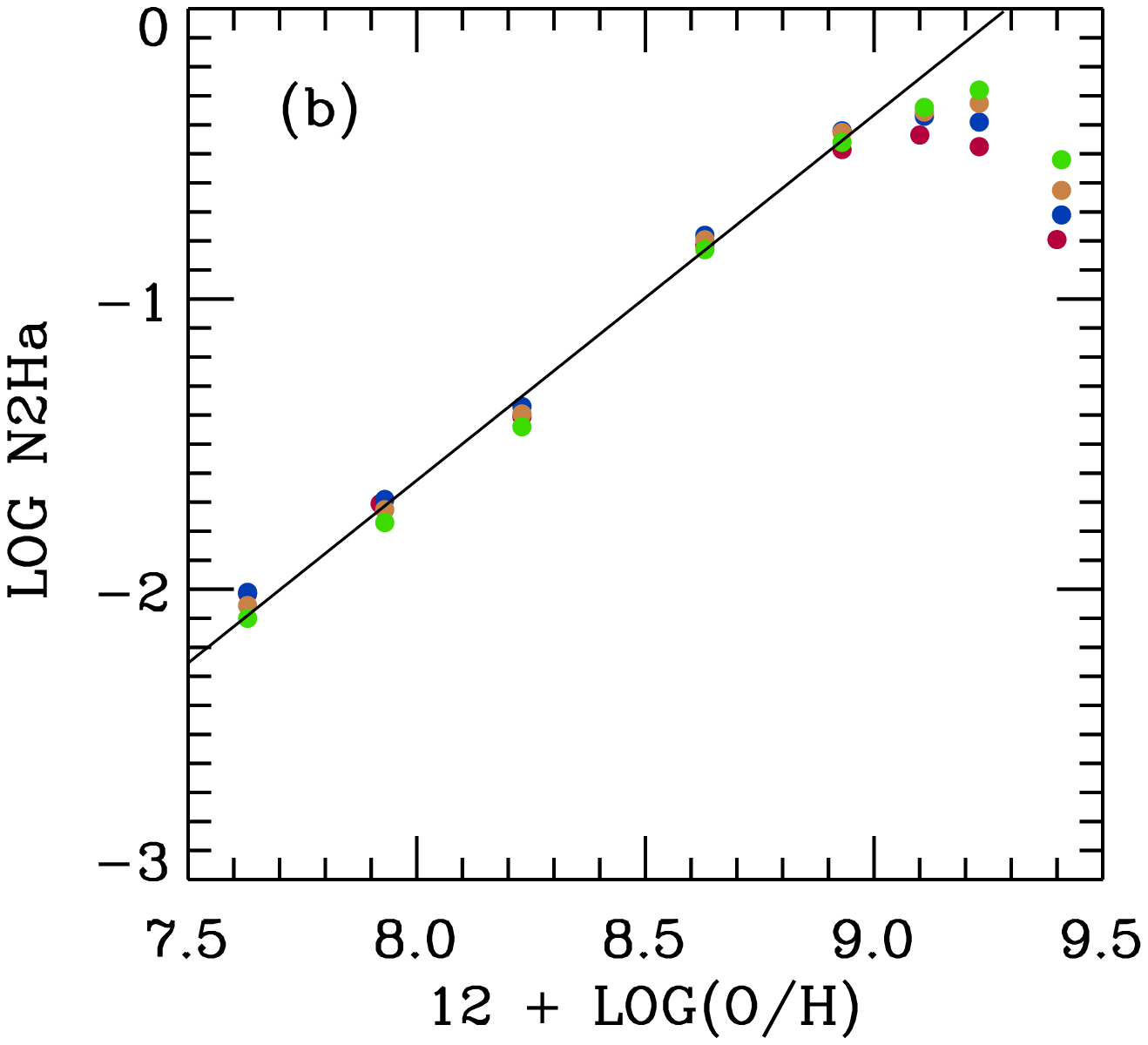}}
\caption{N2Ha  vs.\ 12+$\log$(O/H) as a function of the ionization
parameter, $q$. (a) Model calculations
of \cite{kd02} taken from their Figure~7. The color
of the symbols shows the $\log q$ parameter ranging from
8.18 (green), 7.90 (gold), 7.60 (blue), to 7.30 (red).
(b) Same points from (a) after applying a correction of 
$-0.66(\log q - 7.5)$ 
to NH2a. This brings the points into approximate agreement
throughout much of the $12+\log$(O/H) range.
Below $12+\log$(O/H)=9, the points in (a) are well fit by the linear
relation of Equation~\ref{abundance}, which is shown for
$\log q=7.5$ as the black line in (b).
\label{n2ha_q}
}
\addtolength{\baselineskip}{10pt}
\end{inlinefigure}


\subsection{O32 and R23}
\label{seco32}

The small variation in R23 seen in Figure~\ref{r23_o32}
means that R23 is not 
a useful metallicity diagnostic for either the high-redshift 
sample or the higher-luminosity SDSS sample. However, the
O32 ratio may be a useful metallicity indicator.
\citet{shapley15} argued, using the composite spectra for 
SDSS galaxies from \citet{andrews13}, that there is a strong 
dependence of galaxy metallicity on position in the O32 versus R23 diagram, 
with low-metallicity galaxies lying at high O32 and high-metallicity
galaxies lying at low O32. They therefore suggested that
O32 might be the best approach to measuring the metallicities
of high-redshift galaxies.

\begin{inlinefigure}
\centerline{\includegraphics[angle=0,width=4.0in]{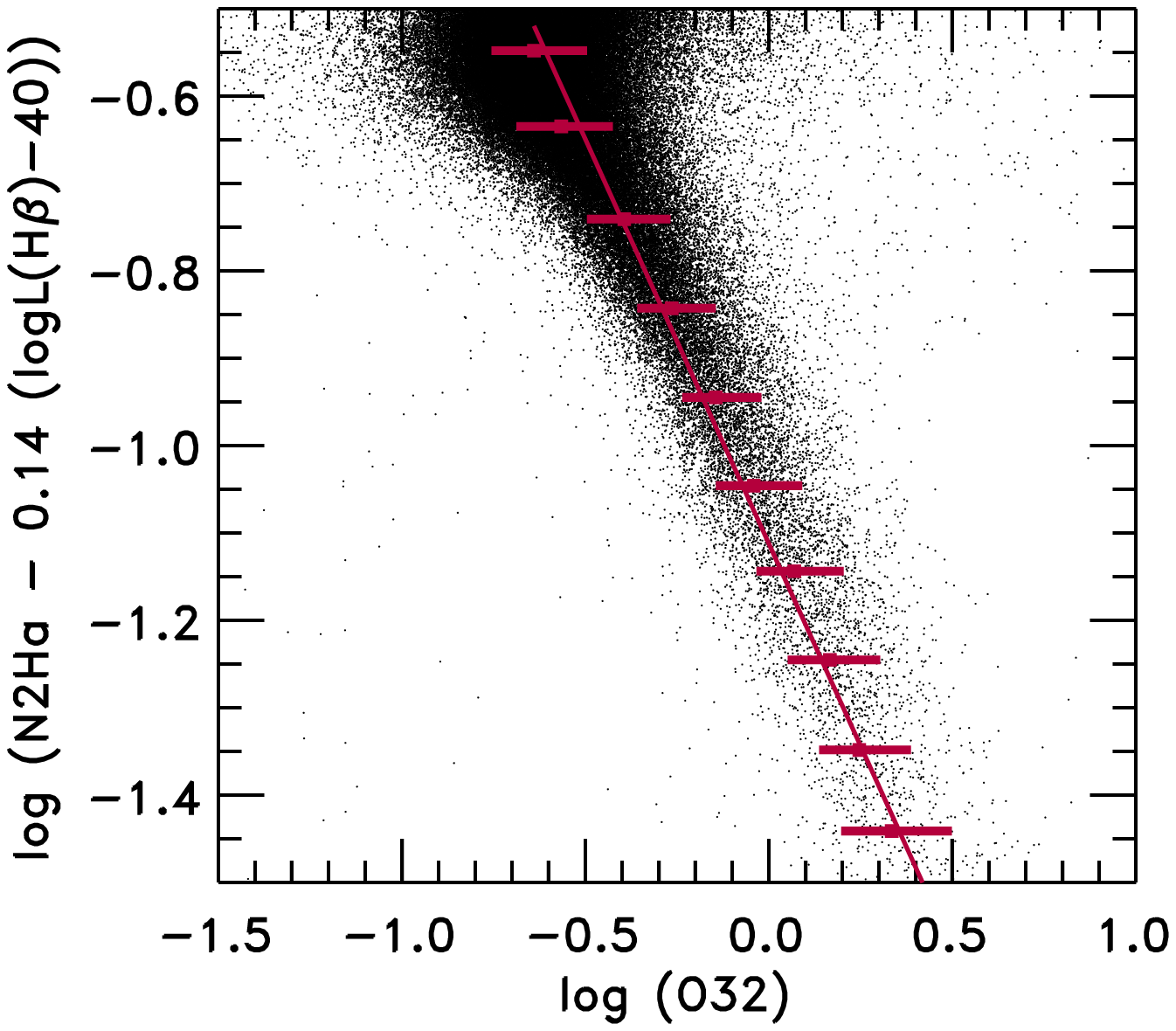}}
\caption{N2Ha$-0.17\log (L({\rm H}\beta$)-40)  vs.\ O32 (black dots)
for the full SDSS sample on the star-forming galaxy locus of the BPT diagram
(N2Ha$\ <-0.5$). The red bars show
the 68$\%$ confidence range in each N2Ha interval, while
the red line shows the linear fit of Equation~\ref{o32_n2ha}.
\label{fig_n2ha_o32}
}
\addtolength{\baselineskip}{10pt}
\end{inlinefigure}

However, O32 is hard to measure because of the wide wavelength range
between the lines, and the necessity for extinction corrections,
which require measurements of both H$\alpha$ and H$\beta$. 
Fortunately, we can avoid the use of O32, because it is quite tightly
correlated with N2Ha for galaxies in the star-forming galaxy locus
of the BPT diagram (N2Ha$<-0.5$). 
We illustrate this in Figure~\ref{fig_n2ha_o32},
where we plot luminosity-adjusted N2Ha versus
O32 for galaxies in the full SDSS sample that satisfy this constraint.
The luminosity adjustment in N2Ha tightens the correlation 
and reduces the dispersion by $\sim20$\%. 
The red bars show the 68\% confidence range in each N2Ha interval,
and the red line shows a fit of the form
\begin{eqnarray}
\log {\rm N2Ha}=-1.11-0.93\log {\rm O32} \nonumber \\
-0.14(\log L({\rm H}\beta)-40) \,.
\label{o32_n2ha}
\end{eqnarray}
Thus, over the full spread in O32, we can use the much more
easily-measured N2Ha as a proxy for O32. A luminosity-adjusted N2Ha
is therefore a promising method for measuring the
the metallicities of high-redshift galaxies.

\section{Summary}
\label{secsummary}

We confirm that all the standard strong emission-line diagnostics
for high-redshift galaxies are shifted relative to the average for SDSS galaxies.
However, the major result of this paper is that these diagnostics are not shifted 
relative to SDSS galaxies at similar values of $L({\rm H}\beta$). 
That is, all galaxies with a given $L({\rm H}\beta$) show invariant emission-line 
properties at all redshifts.

This remarkable result suggests that we can use the 
easier to study SDSS galaxies as proxies for high-redshift  galaxies
of the same luminosity. Through such analyses, we showed that
the increase of N2Ha, for fixed O3Hb, seen in the BPT diagram
as a function of increasing line luminosity is
primarily driven by higher N/O abundances in
higher-line-luminosity galaxies. However,  this effect
is partially offset by an increase in the ionization parameter
with increasing line luminosity which decreases [NII]6584 relative
to [OIII]5007. Changes in N2O2 and N2S2 which increase with
$\log$ L(H$\beta$) are driven solely by the increase in N/O
with increasing luminosity.

Finally, we argue that we can determine the
metallicities of galaxies using a luminosity-adjusted N2Ha parameter
for which we provide the equation (Equation~\ref{ppeqn}).
Such a luminosity-adjusted N2Ha parameter is much easier to 
measure than O32, which also appears to be a useful diagnostic
for higher-line-luminosity galaxies.

\acknowledgements
We thank the referee for very helpful comments
that improved the manuscript. We also
thank Stephanie Juneau and Christy Tremonti
for their valuable input on the first draft of the paper
and Naveen Reddy for productive discussions.
We gratefully acknowledge support from NSF grants
AST-1313309 (L.~L.~C.) and AST-1313150 (A.~J.~B) and
the University of Wisconsin Research Committee with funds
granted by the Wisconsin Alumni Research Foundation (A.~J.~B.).
This research was supported by the Munich Institute for Astro-
and Particle Physics (MIAPP) of the DFG cluster of excellence
``Origin and Structure of the Universe''.


\end{document}